\documentclass[useAMS,usenatbib]{mnras}
  \usepackage{times,graphicx,amssymb,amsmath,natbib}

 \usepackage{hyperref}
\usepackage[pdftex]{}
\pdfoutput=1
\newcommand{\be}{\begin{equation}}
\newcommand{\e}{\end{equation}}
\newcommand{\bear}{\begin{eqnarray}}
\newcommand{\ear}{\end{eqnarray}}

\newcommand{\f}{\frac}
\newcommand{\de}{{\rm d}}

\defcitealias{2015ApJ...813L...8M}{MH15}
\defcitealias{2007ApJ...654..731H}{H07}
\defcitealias{2015A&A...578A..83G}{G15}

\def\prd{Phys. Rev. D}
\def\mnras{MNRAS}
\def\apj{ApJ}
\def\apjl{ApJL}
\def\apjs{ApJS}
\def\aap{A\&A}

\def\araa{ARA\&A}
\def\aj{AJ}

\def\physrep{Phys. Rep.}

\def\pasj{Pub. Astron. Soc. Japan}

\def\nar{New Astronomy Reviews}

\def\HeII{\hbox{He~$\scriptstyle\rm II$}}
\def\HeIII{\hbox{He~$\scriptstyle\rm III$}}
\def\HI{\hbox{H~$\scriptstyle\rm I$}}
\def\HII{\hbox{H~$\scriptstyle\rm II$}}

\begin{document}

\title[Cosmic reionization after Planck II]
{Cosmic reionization after Planck II: contribution from quasars}
\author[Mitra, Choudhury \& Ferrara]
{Sourav Mitra$^1$\thanks{E-mail: hisourav@gmail.com},~
T. Roy Choudhury$^2$\thanks{E-mail: tirth@ncra.tifr.res.in}~
and
Andrea Ferrara$^3$\thanks{E-mail: andrea.ferrara@sns.it}\\
$^1$University of the Western Cape, Bellville, Cape Town 7535, South Africa\\
$^2$National Centre for Radio Astrophysics, TIFR, Post Bag 3, Ganeshkhind, Pune 411007, India\\
$^3$Scuola Normale Superiore, Piazza dei Cavalieri 7, 56126 Pisa, Italy
} 

\maketitle

\date{\today}

\begin{abstract}

In the light of the recent \emph{Planck} downward revision of the electron scattering optical depth,
and of the discovery of a faint AGN population at $z > 4$, we reassess the
actual contribution of quasars to cosmic reionization. To this aim, we extend our  previous MCMC-based
data-constrained semi-analytic reionization model and study the role of quasars on global reionization history.
We find that, the quasars can alone reionize the Universe only for models with very high AGN
emissivities at high redshift. These models are still allowed by the recent CMB
data and most of the observations related to \HI\ reionization. However, they predict
an extended and early \HeII\ reionization ending at $z\gtrsim4$ and a much slower evolution in the
mean \HeII\ Ly-$\alpha$ forest opacity than what the actual observation suggests.
Thus when we further constrain our model against the \HeII\ Ly-$\alpha$ forest data,
this AGN-dominated scenario is found to be clearly ruled out at 2-$\sigma$ limits.
The data seems to favour a standard {\it two-component} picture where quasar contributions
become negligible at $z\gtrsim6$ and a non-zero escape fraction of $\sim10\%$ is needed
from early-epoch galaxies. For such models, mean neutral hydrogen fraction decreases to
$\sim10^{-4}$ at $z=6.2$ from $\sim0.8$ at $z=10.0$ and helium becomes doubly ionized at much
later time, $z\sim3$. We find that, these models are as well in good agreement with the observed
thermal evolution of IGM as opposed to models with very high AGN emissivities.
\end{abstract}

\begin{keywords}
dark ages, reionization, first stars -- intergalactic medium -- quasars: general -- cosmology:
theory -- large-scale structure of Universe.
\end{keywords}

\section{Introduction}

Over the past few decades, various observational studies have been performed in order to
constrain the epoch of hydrogen reionization, among them, observations from high-redshift
QSO spectra \citep{Becker:2001ee,2003AJ....126....1W,2006AJ....132..117F} and CMB by
Wilkinson Microwave Anisotropy Probe (WMAP) \citep{Komatsu:2010fb,2013ApJS..208...20B} or Planck
\citep{2014A&A...571A..16P,2016A&A...594A..13P,2016A&A...596A.107P} have been proven to be most
useful. These observations along with theoretical models and simulations aim to help understand
the nature of the reionization process and the sources responsible for it. Although it is
commonly believed that, the IGM
became ionized by the UV radiation from star forming high redshift galaxies and the bright
active galactic nuclei (AGN) populations, their relative contributions remain poorly
understood \citep{2012ApJ...755..124G,2015A&A...578A..83G,2015ApJ...813L...8M,2016MNRAS.458L..94S,
2016MNRAS.459.2299F}. The rapidly decreasing number density of QSOs and AGNs at $z>3$
\citep{2007ApJ...654..731H} usually leads to the assumption that the high-redshift galaxies
might have dominated hydrogen reionization at $z\gtrsim6$ \citep{BarkanaLoeb01,LoebBarkana01,
2006astro.ph..3360L,tirth06a,tirth09}. Based on this idea, several semi-analytic models have
been put forward using a combination of different observations \citep{tirth05,2005ApJ...625....1W,
2006MNRAS.370.1401G,2007MNRAS.379..253D,2007MNRAS.377..285S,2008MNRAS.391...63I,mitra1,
2011MNRAS.412.2781K,mitra2,mitra3,2014ApJ...785...65C}.

However, recent measurements from Planck brought up an important question regarding the
contribution from high-redshift sources. Unlike the WMAP results, they found relatively small
value for integrated reionization optical depth, $\tau_{\rm el}=0.066\pm0.012$ from Planck 2015;
\citep{2016A&A...594A..13P} or $0.055\pm 0.009$ from Planck 2016; \citep{2016A&A...596A.107P},
which corresponds to a sudden reionization occurring at much lower redshift, mean
$z_{\rm reion}\approx7.8-8.8$ \citep{2016A&A...596A.108P} and thus reduces
the need for high-$z$ galaxies as reionization sources
\citep{2015ApJ...802L..19R,2015ApJ...811..140B,mitra4}. This also seems to explain the
rapid drop observed in the space density of Ly-$\alpha$ emitters at $z\sim 7$
\citep{2015MNRAS.446..566M,2014arXiv1412.4790C,2014ApJ...795...20S}. Using the Planck
constraints along with high-$z$ galaxy luminosity function (LF) data
\citep{2006NewAR..50..152B, 2012ApJ...745..110O,2012ApJ...760..108B,2014ApJ...786..108O,
2014arXiv1411.2976B,2014arXiv1412.1472M,2015ApJ...803...34B}, one can obtain relatively
lower and almost constant escape fraction ($f_{\rm esc}$) of ionizing photons from galaxies
(but also see \citealt{2016arXiv160503970P}),
which was somewhat inevitably higher and evolving towards higher redshift in order to match
the WMAP data \citep{mitra3}. In particular, recently \citet{mitra4} found that a
constant $\sim10\%$ $f_{\rm esc}$ at $z\geqslant6$ is sufficient to ionize the Universe and
can as well explain most of the current observational limits related to \HI\ reionization.
A similar constraint on $f_{\rm esc}$ has been reported by \cite{2015ApJ...811..140B} and
\cite{2016MNRAS.457.4051K}. Likewise, using an improved semi-numerical code, \cite{2016MNRAS.457.1550H}
require $f_{\rm esc} \approx 2-11\%$, independent of halo mass and redshift, to
simultaneously match current key reionization observables. \citet{2015MNRAS.453..960M}
too found a non-evolving and considerably lower ($<5\%$) escape fraction using their high-resolution
cosmological zoom-in simulation for galaxy formation. Such small leakages from
star-forming galaxies again leads to the fact to reconsider their contributions for reionization.

On the other hand, the conventional scenario of rapidly decreasing ionizing QSO background beyond
$z\gtrsim3$ is also changing, owing to various recent claims on finding significantly higher (than
previously expected) number of faint AGNs at higher redshifts ($z>3$) which are able to contribute
a rather steep luminosity function \citep{2011ApJ...728L..26G,2011ApJ...741...91C,2012ApJ...755..124G,
2015A&A...578A..83G}. Interestingly, \citet{2016MNRAS.457.4051K} found that $f_{\rm esc}$ needs to
increase by a factor of $\sim 3$ from $z \sim 3.5$ to 5.5 in order to match the measured photoionization
rates in case these fainter AGNs are not taken into account. As a result, it has been proposed recently
that the reionization may have been driven entirely by the QSOs
\citep{2015ApJ...813L...8M,2016MNRAS.457.4051K}. These models have no difficulty in matching the
$\tau_{\rm el}$ constraints from Planck.
In addition, they lead to two significant features of reionization which might resolve
some puzzling observations of IGM: a late hydrogen reionization completing around $z\sim6$,
which explains the remarkable flatness observed in the \HI\ photoionization rate
\citep{2013MNRAS.436.1023B} and a distinguishing feature of early \HeII\ reionization
ending by $z\sim4$, which seems to be consistent with the measured high-transmission regions
of \HeII\ Ly-$\alpha$ forest out to $z\sim3.5$ \citep{2016ApJ...825..144W}.
However, several recent claims make this AGN-only reionization picture somewhat controversial.
\cite{2015ApJ...813L..35K} claimed that
the number of spectroscopically confirmed faint AGNs at $z\gtrsim6$ might not be high enough to
reionize the Universe alone (see \citealt{2016MNRAS.459.2299F} as well).
Also, \cite{2017MNRAS.468.4691D} pointed out a possible tension between these
AGN-dominated models and observed thermal history of the IGM. 
More recently, based on the SDSS high-redshift quasar sample at $z\sim6$, \cite{2016ApJ...833..222J} 
find that the observed quasar population is not enough to support the AGN-dominated
reionization scenario. Thus, the possibility of a larger
QSO emissivity at higher redshifts combined with the lower $\tau_{\rm el}$ data from Planck
demands more rigorous and careful investigation on the actual contribution from star-forming
galaxies and QSOs to reionization.

In this paper, we extend our previous Markov Chain Monte-Carlo (MCMC)-constrained reionization model with Planck
\citep{mitra4} to explore the possibility of {\it quasar-only} reionization scenario by
comparing different AGN emissivities. The paper is structured as follows.
In the next section, we review the main features of our semi-analytical reionization
model. We consider four different cases to compare their relative quasar contributions.
In Section~\ref{sec:MCMC}, we present the MCMC constraints on reionization scenario obtained from
each model. We compare our predictions with the observation related to the thermal evolution of IGM
in Section~\ref{sec:temperature}. Finally, we summarize and conclude our main findings in Section~\ref{sec:conclusions}.
We assume a flat Universe with Planck cosmological parameters: $\Omega_m = 0.3089$,
$\Omega_{\Lambda} = 1 - \Omega_m$, $\Omega_b h^2 = 0.0223$, $h=0.6774$, $\sigma_8=0.8159$,
$ n_s=0.9667$ and $Y_{\rm P}=0.2453$ \citep{2016A&A...594A..13P}.

\section{Reionization model}
\label{sec:cfmodel}

In the following sections, we develop a robust statistical approach in order to constrain
a simple reionization model against selected datasets. We perform a detailed MCMC analysis over
our model parameters related to the mean free path of ionizing photons and stellar and quasar
emissivities. The method is built on a series of our earlier papers \citep{mitra1,mitra2,mitra3,mitra4}
with the addition of different quasar contribution models. We study their effects on hydrogen
and helium reionization separately and try to see how much information can be gained on reionization
sources by adding helium reionization data in our analysis. Later we also check the consistency
of such models with observed IGM thermal evolution.

Throughout this work, we use a data-constrained semi-analytical reionization model, which is
based on \cite{tirth05} and \cite{tirth06}. Below we summarize its key features.

\begin{itemize}
 \item The model adopts a {\it lognormal} a probability distribution function $P(\Delta)$
 ($\Delta$ being the overdensity of corresponding regions) at low densities, 
 changing to a {\it power-law} at high densities \citep{tirth05}. The ionization and
 thermal histories of all hydrogen and helium regions are evaluated self-consistently
 by treating the IGM as a multi-phase medium. It accounts for the IGM
 inhomogeneities in a very similar manner proposed by \citet{2000ApJ...530....1M},
 where reionization is said to be complete once all the low-density regions with overdensities
 $\Delta < \Delta_{\rm crit}$ are ionized, where the critical density $\Delta_{\rm crit}$
 is determined from the mean separation of the ionizing sources
 \citep{2000ApJ...530....1M,2003ApJ...586..693W}.
 According to this prescription, the mean free path of photons is
 computed from the distribution of high density regions:
\be
\lambda_{\rm mfp}(z) = \f{\lambda_0}{[1 - F_V(z)]^{2/3}}
\label{eq:lambda_0}
\e
where $F_V$ is the volume fraction of ionized regions, i.e., 
 \be
  F_V(\Delta_i)=\int_0^{\Delta_i} {\rm d}\Delta P(\Delta)
\label{eq:F_V}
 \e
and $\lambda_0$ is a normalization parameter which we take as a free parameter.
We assume that, $\Delta_i$ does not evolve significantly with
time in the pre-overlap stage; it is equal to a critical value $\Delta_{\rm crit}$.
In this work, it has been taken as a free parameter. 
Once $\Delta_{\rm crit}$
is known, we follow the evolution of the ionized volume filling factor
of the corresponding regions, $Q_i$, until it becomes unity and following that,
we enter the post-overlap stage.

\item {\it Stellar contribution}: Reionization is assumed to be driven by PopII
stellar sources which have sub-solar metallicities and a Salpeter IMF in the
mass range $1 - 100 M_{\odot}$. We neglect the contributions from the metal-free
PopIII sources as the impact of those on reionization is likely to be insignificant
due to low electron scattering optical depth data from Planck \citep{mitra4}.
The model also calculates the suppression of star formation in low-mass haloes
i.e. {\it radiative feedback} self-consistently from the thermal evolution of
IGM (\citealt{tirth05}; also see \citealt{2008MNRAS.390..920O,2013MNRAS.432.3340S}).
The production rate of ionizing photons in the IGM is then computed as
 \be
  \dot{n}_{\rm ph}(z) = n_b N_{\rm ion} \f{\de f_{\rm coll}}{\de t} 
  \equiv n_b \epsilon_{\rm II} N_{\gamma} \f{\de f_{\rm coll}}{\de t}
 \e
where, $f_{\rm coll}$ is the collapsed fraction of dark matter haloes, $n_b$ is
the total baryonic number density in the IGM, $N_{\rm ion}$ is the number of
photons entering the IGM per baryon in stars and $\epsilon_{\rm II}$ is the
product of the star formation efficiency $\epsilon_*$ and escape fraction
$f_{\rm esc}$ of the ionizing photons from stars. We treat $\epsilon_{\rm II}$ as a free parameter
in this analysis.
\end{itemize}

\subsection{Comoving QSO emissivity}
\label{sec:qso_emissivity}

The model also incorporates the contribution of quasars by computing their
ionizing emissivities. In our earlier works \citep{mitra1,mitra2,mitra4},
we adopted a fixed model for the AGN contribution and calculated the number of ionizing
photons from QSOs by integrating their observed luminosity functions  at $z<6$
(\citealt{2007ApJ...654..731H}; hereafter, \citetalias{2007ApJ...654..731H}),
where we assumed that they have negligible effects on IGM at higher redshifts
\citep{tirth05}. The comoving QSO emissivity $\epsilon_\nu$ at $912 \mathring{\rm A}$
($\epsilon_{912}^{\rm H07}$ in units of ${\rm erg~s^{-1}~Hz^{-1}~Mpc^{-3}}$)
is then obtained by integrating the QSO luminosity function at each redshift assuming
its efficiency to be unity (but also see \citealt{2009ApJ...703.1416F}), which is
identical to integrating the B-band luminosity functions with the conversion
$L_{\nu}(912 \mathring{\rm A}) = 10^{18.05} {\rm erg~s^{-1}~Hz^{-1}}(L_B/L_\odot)$
\citep{2003ApJ...584..110S,tirth05,2006ApJS..166..470R,2007ApJ...654..731H}.
However, this standard picture of negligible QSO contributions at $z>3$
has been challenged by the recent findings of numerous faint AGN candidates
at higher redshifts \citep{2015A&A...578A..83G}. Based on this, \cite{2015ApJ...813L...8M}
proposed an AGN comoving emissivity fit which is significantly higher at high-$z$.
The faint AGN population alone can dominate the reionization process in their model.
We shall refer this as \citetalias{2015ApJ...813L...8M} model.

However, it has been argued that the observations of faint end QSO LFs at high-$z$ ($z>3$)
are very uncertain \citep{2016arXiv160204407Y}. So, rather than taking a fixed
AGN model, in this work we aim to get the constraints on high-redshift LF slope
backward by varying it as an additional free parameter. This will help us to gain
more insight on the relative contributions of the stellar and AGN components allowed
by the current data. For that, we take a model with comoving AGN emissivity same as
that provided by \citetalias{2007ApJ...654..731H} model for $z\le2$, while its
redshift evolution at $z>2$ is accompanied by an exponential fall depending on the
free parameter $\beta$, so that this $\beta$ will now determine the high-$z$ slope:
\begin{align}
 \epsilon_{912}(z) &= \epsilon_{912}^{\rm H07}(z), & \text{for } z\le2\nonumber\\
                   &= \epsilon_{912}^{\rm H07}(z=2)\times e^{\beta\left(z-2\right)^2}, & \text{for } z>2
\end{align}
This method is quite similar to what described in the recent study by
\cite{2016arXiv160204407Y}, where they try to constrain the contribution
of high-$z$  galaxies and AGNs to reionization by varying $f_{\rm esc}$
and $\beta$ (although the definition of $\beta$ is different in their case)
while simultaneously satisfying the Planck electron scattering optical depth
data and the redshift for completion of hydrogen or helium reionization.
However, note that our analysis, in principle, should give a more robust
constraint on these parameters since it is also constrained by many other
observables. We assume the quasars have
a frequency spectrum $\epsilon_{\nu} \propto \nu^{-1.57}$ for frequencies
above the hydrogen ionization edge.

The above parametrization of the quasar emissivity can fairly reproduce
both the \citetalias{2007ApJ...654..731H}
($\beta=-0.27$) and \citetalias{2015ApJ...813L...8M} ($\beta=-0.01$)
cases\footnote{From here on, \citetalias{2007ApJ...654..731H} and
\citetalias{2015ApJ...813L...8M} refer to the models with
$\beta=-0.27$ and $\beta=-0.01$ respectively.}
(see Fig.~\ref{fig:emissivity}). Note that, at lower redshifts ($z\lesssim4$)
the comoving AGN emissivity for our $\beta=-0.01$ model is lower than the
original \citetalias{2015ApJ...813L...8M} fit. In that sense, we somewhat
underestimate the AGN contributions in compare to their model at those redshifts.
But we shall see that our overall conclusions will remain unaffected by this
reconciliation. Our main goal is to investigate whether a faint AGN
population alone can dominate the reionization process, or
equivalently find out what are the acceptable ranges of $\beta$ allowed by current
observations related to reionization. We hence treat $\beta$ as a free
parameter along with our other model parameters.

\subsection{Free parameters and datasets}
\label{sec:hyd_data}

Although, the above model can be constrained by a variety of observational datasets,
we check that most of the relevant constraints come from a subset of those data
\citep{mitra1}. So, to keep this analysis simple we obtain the constraints on
reionization using {\it mainly} three datasets:
\begin{enumerate}
\item observed HI Ly$\alpha$ effective optical depth $\tau_{\rm eff, HI}$ at $2.4\leqslant z\leqslant6$,
as obtained from the QSO absorption spectra. The values adopted for $\tau_{\rm eff, HI}$ in this paper are
based on the data tabulated in \cite{2006AJ....132..117F} and \cite{2013MNRAS.430.2067B}. We bin the data
into redshift bins of width $\Delta z=0.5$ and compute the mean in each bin. The corresponding uncertainties
are estimated using the extreme values of $\tau_{\rm eff, HI}$ along different lines of
sight\footnote{In our earlier works \citep{mitra1,mitra2,mitra3} we have used the hydrogen photoionization
rate measurements \citep{2007MNRAS.382..325B,2013MNRAS.436.1023B} instead of the $\tau_{\rm eff, HI}$
used in this work. We have checked and found that the main conclusions of the paper remain unchanged
irrespective of which of the two data sets is used in the analysis.};
\item redshift evolution of Lyman-limit systems (LLS), $\de N_{\rm LL}/\de z$ over a wide redshift
range $0.36 < z < 6$ from the combined data of \cite{2010ApJ...721.1448S} and \cite{2010ApJ...718..392P}
with the errors calculated using the quadrature method;
\item Thomson scattering optical depth $\tau_{\rm el}$ data ($0.055\pm0.009$) from Planck 2016
\citep{2016A&A...596A.107P}.
\end{enumerate}
We also impose somewhat model-independent upper limits on the neutral hydrogen fraction $x_{\rm HI}$
at $z\sim5-6$ from \cite{2015MNRAS.447..499M} as a prior to our model.

As mentioned earlier, the free parameters of this model are $\epsilon_{\rm II}$, $\lambda_0$,
$\Delta_{\rm crit}$ and $\beta$; all the cosmological parameters are fixed at
their best-fit Planck value. In principle, $\epsilon_{\rm II}$ can have a dependence on redshift
$z$ and halo mass $M$, but for simplicity, we assume $\epsilon_{\rm II}$ to be independent of $z$ or $M$
throughout this work. This is also motivated by the results from our earlier work \citep{mitra4},
where we find that both $f_{\rm esc}$ and $\epsilon_*$ are almost non-evolving with redshift.
Unlike our previous works \citep{mitra1,mitra2,mitra3}, we allow $\Delta_{\rm crit}$ to be a free
parameter. In our models $\Delta_{\rm crit}$ sets the mean free path of ionizing photons and its
value depends on the typical separation between the ionizing sources
\citep{2000ApJ...530....1M}. Since QSOs are relatively
rarer sources, the implied $\Delta_{\rm crit}$ is expected to be relatively higher for
QSO-dominated models than for the stellar-dominated ones. For our simplified mean free path
prescription, this value usually turns out to be $\sim 20-60$ depending on the density profile
of the halo \citep{tirth05,tirth09}. Keeping this in mind, we allow it to vary only in the
range $[20, 60]$. We also impose a prior $\beta \leq 0$ to ensure that
the quasar emissivities do not diverge at high redshifts.

\subsection{Inclusion of helium reionization data}
\label{sec:he_data}

As we will see in the following section(s), the constraints on the QSO emissivity models are
relatively weaker when we consider only those datasets related to the hydrogen reionization.
The AGNs can produce significantly more hard ionizing photons compared to stellar sources and
should ionize \HeII\ more efficiently. Thus the helium reionization history should have a more
decisive dependence on $\beta$ than the \HI\ reionization \citep{2016arXiv160204407Y}. 
Observations of the \HeII\ Ly$\alpha$ forest have already been used to study the
\HeII\ reionization \citep{2005MNRAS.361.1399G,2009ApJ...706..970D,2009ApJ...694..842M,
2013MNRAS.431L..53K}. 

Given this, we include one more observable related to helium reionization,
namely the effective optical depth of \HeII\ Ly$\alpha$ absorption $\tau_{\rm eff, HeII}$
data at $2.3<z<3.5$ from \cite{2016ApJ...825..144W}. This observation manifests surprisingly
low \HeII\ absorption at $z\sim3.5$, which might indicate that the bulk of the helium was
ionized much earlier (at $z>3.5$). This is somewhat in tension with the general findings from
current numerical radiative transfer simulations \citep{2009ApJ...694..842M,2013MNRAS.435.3169C,
2014MNRAS.445.4186C}, where the \HeII\ reionization ends at relatively later epoch,
$z\approx3$ \citep{2016ARA&A..54..313M,2017MNRAS.468.4691D}. Thus, it would be interesting to
see what sources could support such an early reionization within our semi-analytic formalism.

The observed values of $\tau_{\rm eff, HeII}$ along different sightlines
show a large scatter particularly at $z \gtrsim 3$ \citep{2016ApJ...825..144W}. In order to adapt
them into our likelihood analysis, we have binned the data points within redshift intervals of
$z=0.2$ and calculated the mean. The errors are calculated using the extreme values of
$\tau_{\rm eff, HeII}$ along different lines of sight. 

The inclusion of the $\tau_{\rm eff, HeII}$ would imply additional calculations in our
theoretical model. We briefly summarize the additional steps which have been included
for this purpose, the details can be found in \citet{tirth05}.

\begin{itemize}

\item As in the case of hydrogen reionization, we assume the \HeII\ in the low-density
$\Delta < \Delta_{\rm crit, HeII}$ to be ionized first, where $\Delta_{\rm crit, HeII}$
is the critical overdensity similar to the $\Delta_{\rm crit}$ used for hydrogen reionization.
It is essentially determined by the mean separation between helium ionizing sources and
can, in principle, be different from $\Delta_{\rm crit}$. However, we find that our results
are relatively insensitive to the exact chosen value of $\Delta_{\rm crit, HeII}$, hence we
avoid introduction of one more free parameter in our model and simply use
$\Delta_{\rm crit, HeII} = \Delta_{\rm crit}$.

\item We use relations analogous to equations (\ref{eq:lambda_0}) and (\ref{eq:F_V}) to
estimate the mean free path $\lambda_{\rm mfp, HeII}$ for \HeII\ ionizing photons. Given the
emissivity $\epsilon_{\nu}$ and $\lambda_{\rm mfp, HeII}$, it is straightforward to calculate
the photoionization rate for \HeII\ (assuming $\lambda_{\rm mfp, HeII}$ to be much smaller than
the horizon size, which holds true for $z \gtrsim 2$).

\item We evolve the average fraction of different ionization states of helium, along with that
of hydrogen and the temperature, which can then be used for calculating the Ly$\alpha$ optical
depth $\tau_{\rm HeII}$ for helium. The effective optical depth is simply obtained from the
average value of the corresponding transmitted flux and is given by
\be
\tau_{\rm eff, HeII} = -\ln \left\langle {\rm e}^{- \tau_{\rm HeII}} \right\rangle.
\e

\end{itemize}

\section{Results: MCMC constraints}
\label{sec:MCMC}
We perform an MCMC analysis over all the parameter space 
\{$\epsilon_{\rm II},\lambda_0,\Delta_{\rm crit}, \beta$\} using the above mentioned datasets.
We employ a code based on the publicly available COSMOMC \citep{2002PhRvD..66j3511L}
code and run a number of separate chains until the usual Gelman and Rubin convergence
criterion is satisfied. This method based on the MCMC analysis has already been developed
in our previous works \citep{mitra1,mitra2,mitra4}.

While presenting the results, we clearly distinguish between two cases:
(i) ``without \HeII\ data'' where we obtain constraints using only hydrogen reionization data
(Section \ref{sec:hyd_data}), and (ii) ``with \HeII\ data'' where we include the
$\tau_{\rm eff, HeII}$ data as well (Section \ref{sec:he_data}). This is done to emphasize
the significance of the \HeII\ reionization data in constraining the QSO emissivity models.

\begin{table*}
\begin{tabular*}{0.75\textwidth}{@{\extracolsep{\fill} } l c c | c c }
Parameters & \multicolumn{2}{c}{best-fit with 95\% C.L.} & \multicolumn{2}{c}{fixed $\beta$-model (No MCMC)}\\
& without \HeII\ data & with \HeII\ data & H07 ($\beta=-0.27$) & MH15 ($\beta=-0.01$)\\
\hline
$\beta$ & $-0.06~~[-3.92, 0.0]$ & $-0.43~~[-0.99, -0.19]$ & $-0.27$ & $-0.01$\\
$\epsilon_{\rm II} \times 10^{3}$ & $2.51~~[0.40, 4.94]$ & $3.52~~[2.47, 4.56]$ & $3.30$ & $1.50$\\
$\lambda_0$         & $4.13~~[1.68, 6.71]$ & $3.35~~[1.99, 7.18]$ & $3.85$ & $4.96$\\
$\Delta_{\rm crit}$ & $58.14~~[25.32, 59.97]$ & $52.63~~[25.63, 59.96]$ & $60.0$ & $60.0$\\
\hline
$\tau_{\rm el}$     & $0.058~~[0.056, 0.070]$ & $0.062~~[0.058, 0.070]$ & $0.062$ & $0.061$\\
\hline
\end{tabular*}
\caption{Best-fit value and 95\% C.L. errors of the model parameters (above four) and derived parameter
  (bottom row) obtained from the current MCMC analysis for two cases: with and without \HeII\ data.
  We also show the \citetalias{2007ApJ...654..731H} and \citetalias{2015ApJ...813L...8M} models with
  $\beta=-0.27$ and $-0.01$ respectively, where we fix $\Delta_{\rm crit}=60$ and choose some $\epsilon_{\rm II}$
  and $\lambda_0$ so that they can fairly match all the datasets considered here for hydrogen reionization.}
\label{tab:MCMC}
\end{table*}

\begin{figure}
\centering
  \includegraphics[height=0.38\textwidth, angle=0]{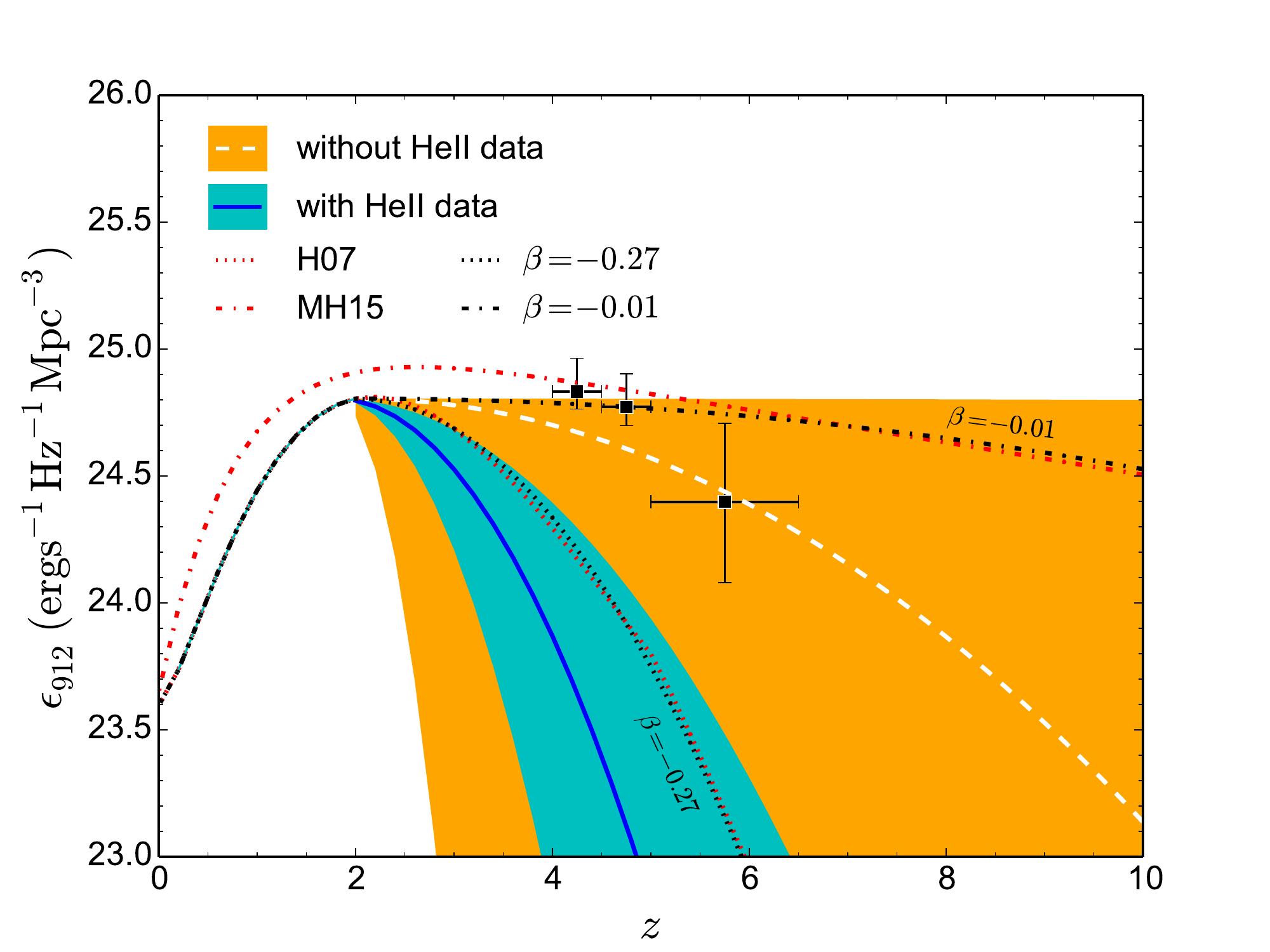}
  \caption{MCMC constraints on ionizing comoving AGN emissivity for two different cases:
  (i) without taking \HeII\ data - white dashed lines for best-fit result with
  orange shaded region around it for 2-$\sigma$ C.L., (ii) with \HeII\ data -
  solid blue curve for the best-fit with shaded cyan region for 2-$\sigma$ limits.
  The \citetalias{2007ApJ...654..731H} (or $\beta=-0.27$) and
  \citetalias{2015ApJ...813L...8M} (or $\beta=-0.01$) models are also shown here
  by dotted and dot-dashed curves respectively.The black points with errorbars are
  the observed data inferred from QLFs by \citet{2015A&A...578A..83G}.}
\label{fig:emissivity}
\end{figure}

\begin{figure*}
\centering
  \includegraphics[height=0.502\textwidth, angle=0]{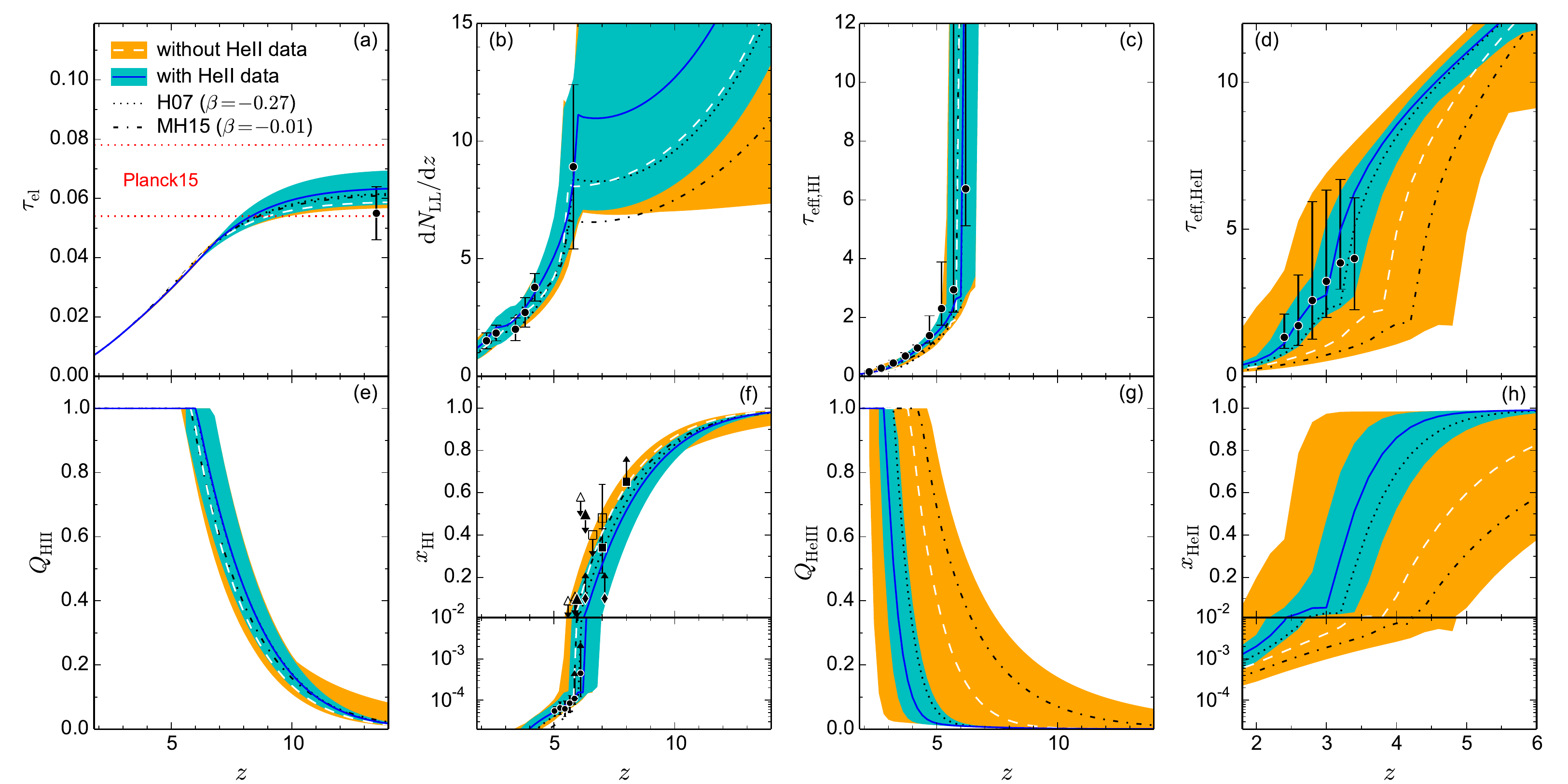}
  \caption{The MCMC constraints on various quantities related to reionization.
  Different panels indicate: (a) electron scattering optical depth $\tau_{\rm el}$,
  (b) redshift evolution of Lyman-limit systems $\de N_{\rm LL}/\de z$,
  (c) effective \HI\ Ly$\alpha$ optical depth $\tau_{\rm eff, HI}$,
  (d) effective \HeII\ Ly$\alpha$ optical depth $\tau_{\rm eff, HeII}$,
  (e) volume filling factor of ionized hydrogen $Q_{\rm HII}$,
  (f) neutral hydrogen fraction $x_{\rm HI}(z)$,
  (g) \HeIII\ ionized volume filling factor $Q_{\rm HeIII}$ and
  (h) global \HeII\ fraction $x_{\rm HeII}$.
  The solid blue and dashed white lines correspond to
  the mean evolution obtained from the models with \HeII\ data and without \HeII\ data
  respectively. The shaded regions refer to the 2-$\sigma$ confidence limits around the mean value
  of the corresponding models. The black points with errorbars are the current
  observational limits on reionization, see the main text for references. For comparison, we
  also plot the \citetalias{2007ApJ...654..731H} (i.e. $\beta=-0.27$) and
  \citetalias{2015ApJ...813L...8M} ($\beta=-0.01$) models.}
\label{fig:MCMC1}
\end{figure*}

\begin{figure}
\centering
  \includegraphics[height=0.38\textwidth, angle=0]{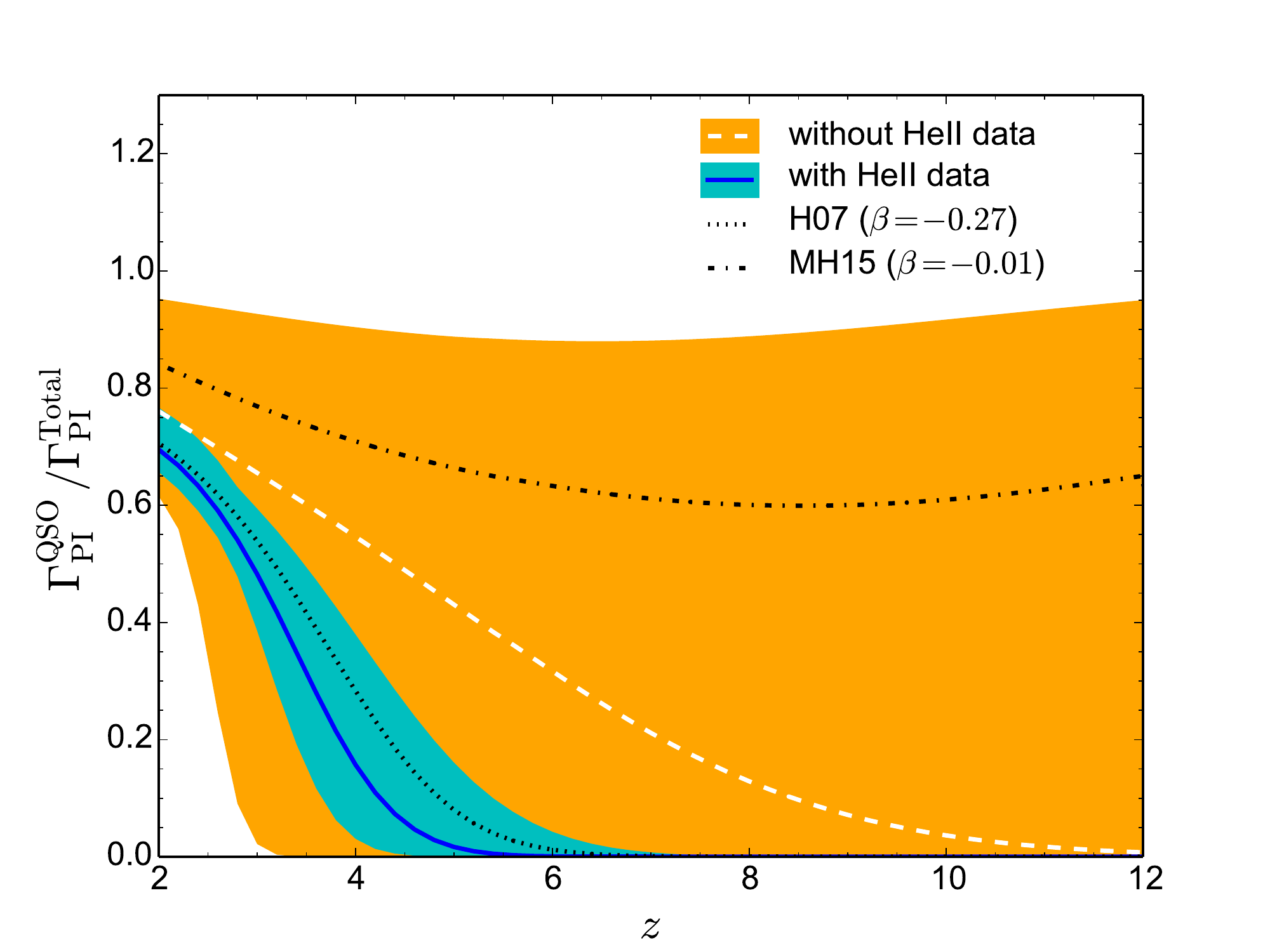}
  \caption{Same as Fig.~\ref{fig:MCMC1}, but showing the constraints on the ratio
  of photoionization rates coming from quasars to its total (galaxies~+~quasars) value
  at different redshifts.}
\label{fig:Gamma_ratio}
\end{figure}

The results from our current MCMC analysis for these two models are summarized in
Table~\ref{tab:MCMC}, where the first four rows are for the free parameters of the model and the last
one ($\tau_{\rm el}$) corresponds to the derived parameter. For comparison, we
also show here \citetalias{2007ApJ...654..731H} ($\beta=-0.27$) and \citetalias{2015ApJ...813L...8M}
($\beta=-0.01$) cases. For that, we fix the $\Delta_{\rm crit}=60$ (following our earlier works)
and choose some combination of $\epsilon_{\rm II}$ and $\lambda_0$ (without running the MCMC)
so that they can reasonably match all the observed datasets for HI Ly$\alpha$ effective optical depth,
redshift evolution of LLSs and also produce $\tau_{\rm el}$ which is allowed by Planck 2016.
We have plotted the MCMC constraints on the comoving AGN emissivities in Fig.~\ref{fig:emissivity}
and the other quantities of interest in Fig.~\ref{fig:MCMC1}. The constraints on the fraction of
hydrogen photoionization rates contributed by QSOs (compared to its total, i.e., QSOs + galaxies, value)
are shown in Fig.~\ref{fig:Gamma_ratio}. For comparison, we also show the
\citetalias{2007ApJ...654..731H} and \citetalias{2015ApJ...813L...8M} models in these three figures.

\subsection{Constraints ``without \HeII\ data''}

The 2-$\sigma$ or 95\% confidence limits on our model parameters without \HeII\ data are given
in the second column of Table~\ref{tab:MCMC} and shown by the shaded orange regions in
Figs~\ref{fig:emissivity}--\ref{fig:Gamma_ratio}. The best-fit model is shown by the dashed
white curves in these figures. We can see from the table that the data allows a wide range
of $\beta$ values spanning from $\sim-4$ (signifying that the quasar contributions are almost
negligible at $z>2$) to $\sim 0$ (signifying that the quasars dominate even at higher redshifts
and a very little contribution is coming from galaxies). This is also evident from
Fig.~\ref{fig:emissivity} where we see that the shaded orange region is remarkably wide.
Both the \citetalias{2007ApJ...654..731H} (dotted black curve) and \citetalias{2015ApJ...813L...8M}
(dot-dashed black curve) cases are well inside this allowed region at $z \gtrsim 4$. 

From the MCMC constraints on various quantities related to reionization shown in Fig.~\ref{fig:MCMC1},
we find that the allowed 2-$\sigma$ confidence limits are relatively narrower for low-$z$ regime
and increase at $z\gtrsim6$. This is expected as most of the datasets used in this work exist only
at low redshifts $z \leq 6$, whereas the higher-$z$ epoch is poorly constrained \citep{mitra1,mitra4}.
The reionization optical depth for the best-fit model is in a good agreement with the Planck 2016 data
(black point with errorbar in panel a). We also show the earlier Planck 2015 limit by dotted red lines.
In panel b, we plot the evolution of Lyman-limit systems which again matches
he combined observational data points from \cite{2010ApJ...721.1448S} and \cite{2010ApJ...718..392P}
at $2<z<6$ quite reasonably for the constraints obtained.
We plot the evolution of effective optical depth of \HI\ Ly$\alpha$ absorption $\tau_{\rm eff, HI}$
in panel c. The MCMC constraint obtained is in general agreement with the observational
measurements from \cite{2006AJ....132..117F,2013MNRAS.430.2067B} in the interval $2 < z < 6$.
From the evolution of volume filling factor $Q_{\rm HII}$ of \HII\ regions (panel e) and the global
neutral hydrogen fraction $x_{\rm HI}$ (panel f), we find that the hydrogen reionization history is
reasonably  well constrained, in spite of the quasar emissivity being allowed to take such wide
range of values. This is because any variation in $\beta$ is appropriately compensated by a similar
change in $\epsilon_{\rm II}$ making sure that the models agree with the observations. In fact, the
allowed range of $\epsilon_{\rm II}$, as can be seen from Table~\ref{tab:MCMC}, is also quite wide
between $\sim 4 \times 10^{-4} - \sim 5 \times 10^{-3}$. 
We also find that the model can match the current observed constraints on $x_{\rm HI}$ quite reasonably
within their errorbars. Note that the match is quite impressive, given the fact that we did
not include these datasets, except the \cite{2015MNRAS.447..499M} data at $z\sim5-6$, as constraints
in the current analysis. 
The observational limits (black points) are taken from various
measurements by
  \cite{2006AJ....132..117F} (filled circle), \cite{2015MNRAS.447..499M} (open triangle), 
  \cite{2006PASJ...58..485T,2013ApJ...774...26C} (filled triangle),
  \cite{2011MNRAS.416L..70B,2013MNRAS.428.3058S} (filled diamond),
  \cite{2008ApJ...677...12O,2010ApJ...723..869O} (open square),
  \cite{2014ApJ...795...20S} (filled square).

Moving on to quantities related to \HeII\ reionization, we plot the effective optical depth
$\tau_{\rm eff, HeII}$ in panel d. The points with error bars are from \citet{2016ApJ...825..144W},
binned appropriately for the MCMC analysis. Note that, these data points are \emph{not taken into account}
for the results presented in this subsection. Clearly, ignoring these data points leads to the 2-$\sigma$
allowed regions that are considerably wide which is also related to the fact that $\beta$ is allowed
to take a wide range of values. We show the evolution of \HeIII\ ionized volume fraction $Q_{\rm HeIII}$
and the global \HeII\ fraction $x_{\rm HeII}$ in panels g and h, respectively. The case $Q_{\rm HeIII}=1$
implies that helium is doubly reionized, and the $x_{\rm HeII}$ in that case is determined by the residual
\HeII\ fraction in the \HeIII\ regions. The global \HeII\ reionization history mainly depends on the
contribution from QSOs and thus leaves a distinguishing feature for different $\beta$ models - higher the value
of $\beta$, earlier the \HeII\ reionization occurs. The best-fit model without
\HeII\ data or the AGN-dominated \citetalias{2015ApJ...813L...8M} model indicates that,
the helium becomes doubly reionized at $z\gtrsim4$, however, the 2-$\sigma$ limits allow a wide range of
reionization redshifts. For example, \HeII\ reionization is allowed to be completed as early as $z \sim 5$
(implying a high $\beta \sim 0$) and as late as $z \sim 2.5$ (implying a low $\beta \sim -4$).

From Fig.~\ref{fig:Gamma_ratio}, we find that the fraction of the photoionization rate contributed by the
quasars can take any value between $0$ and $\sim 90\%$ at $z > 3$. Interestingly, the data allow the
fraction to have consistently high value $\sim 90\%$ over the full redshift range $2 < z < 12$
considered here. These models would correspond to quasar dominated reionization scenarios which,
as we can see, cannot be ruled out by hydrogen reionization data alone.

\subsection{Constraints ``with \HeII\ data''}

The constraints on the model parameters change significantly when we include the $\tau_{\rm eff, HeII}$
data in our analysis. The 2-$\sigma$ or 95\% confidence limits on our model parameters with \HeII\ data
are given in the third column of Table~\ref{tab:MCMC} and shown by the shaded cyan regions
in Figs~\ref{fig:emissivity}--\ref{fig:Gamma_ratio}. The solid blue curves in these figures correspond
to the best-fit model. The values quoted in the table show that the confidence limits on $\beta$ and
$\epsilon_{\rm II}$ are considerably shrunk when the \HeII\ data are accounted for. The allowed values
of $\beta$ are between $-0.99$ and $-0.19$, which accommodates the \citetalias{2007ApJ...654..731H}
($\beta = -0.27$) model, but strongly disfavours the \citetalias{2015ApJ...813L...8M} ($\beta = -0.01$) model.
This can also be seen from Fig \ref{fig:emissivity}. In fact, the data now favours a considerably
lower emissivity at $z\sim4-5$  and is only marginally consistent with the value inferred from QLFs
observation by \cite{2015A&A...578A..83G} (black points with errorbars).

One can see from Table~\ref{tab:MCMC} that, the constraints with \HeII\ data require a
significant contribution from the stars as reionization sources $\epsilon_{\rm II}\sim0.003-0.005$
(2-$\sigma$ limits), which in turn corresponds to an escape fraction $\sim8\%-13\%$  at $z\gtrsim6$
of ionizing photons from galaxies. This is consistent with the results from our earlier work \citep{mitra4},
where we used the \citetalias{2007ApJ...654..731H} model. 

From the MCMC constraints shown in  Fig.~\ref{fig:MCMC1}, we find, as expected, that the effect of including
\HeII\ data on quantities related to hydrogen reionization is negligible (panels a, b, c, e, f). The best-fit
model with \HeII\ data supports a slightly earlier hydrogen reionization epoch; mean $x_{\rm HI}$ goes
from $\sim0.8$ to $\sim10^{-4}$ between $z=10.0$ and $z=6.2$. However, the constraints on \HeII\ reionization
are significantly different compared to the earlier case. This is solely driven by the $\tau_{\rm eff, HeII}$
data (panel d) which essentially disfavours a large set of models which were otherwise allowed by the hydrogen
reionization datasets. We find that the best-fit effective \HeII\ Ly$\alpha$ opacity
increases rapidly from $z=2.4$ to $z=3.4$ by a factor of $\sim6$. The data puts a tight constraint on the
\HeII\ reionization redshift between $2.5-3.5$, with the best-fit reionization redshift being $z = 3$.
The model agrees quite well with the interpretation of the \HeII\ Ly$\alpha$ opacity data by \citet{2016ApJ...825..144W}
who suggested that helium is predominantly in the doubly ionized state in $\sim 50\% ~(\sim 90\%)$ of the IGM
at $z \simeq 3.4$  $(z \simeq 2.8)$. This should be compared with our predictions of the volume filling factor
of \HeIII\ regions for the best-fit model which gives $Q_{\rm HeIII} \sim 50\%$ at $z\simeq3.4$, remarkably
consistent with that given by \citet{2016ApJ...825..144W}. The corresponding value at
$z \simeq 2.8$ is $Q_{\rm HeIII} \sim 95\%$ for our model which is slightly higher than
the \citet{2016ApJ...825..144W} value, however, the two results are fully consistent
if we account for the statistical uncertainties in our analysis.
The \citetalias{2007ApJ...654..731H} model also yields a very similar evolution
as the best-fit blue curve. Models with $\beta<-1 (\beta>-0.2)$ imply a very late (early) reionization
history and fail to match observational data.  However, an improved survey with much larger sample
of $z > 3$ \HeII\ sightlines will be needed to put a tighter constraint on $\beta$ in future.

The fraction of photoionization rate contributed by the quasars, as shown in Fig~\ref{fig:Gamma_ratio},
is significantly small. It can take values at most $\sim 10\%$ ($\sim 50\%$) at $z = 6 (4)$,
thus implying that quasar dominated reionization models at $z > 6$ are strongly disfavoured.
For the best-fit model with \HeII\ data and \citetalias{2007ApJ...654..731H}, the AGN contributions
become negligible at $z\gtrsim6$, which is in stark contrast against the \citetalias{2015ApJ...813L...8M}
case where the AGNs dominate ($\gtrsim60\%$ of total value) even at higher redshifts.

As far as the other quantities are concerned, we find that the best-fit value of $\Delta_{\rm crit}$
is slightly larger for the model without \HeII\ data, as it allows higher contribution from the QSOs
than the model with \HeII\ data. However, both the models support a much broader range for this parameter;
$\Delta_{\rm crit}\sim25-60$ within the 2-$\sigma$ limits. Thus, our results do not vary considerably for
the choice of $\Delta_{\rm crit}$ as long as it is within this limit. We also find that, the electron
scattering optical depths $\tau_{\rm el}$ for all the best-fit models are in good agreement with the
Planck 2016 value. Interestingly, when we include \HeII\ data, the best-fit model predicts slightly
higher $\tau_{\rm el}$ than the other models, as it allows the highest stellar contributions among them.

Interestingly, we find that for the best-fit \HeII\ model which is consistent with all the datasets
considered in this paper, the stellar component of the comoving
ionizing emissivity (in units of s$^{-1}$ Mpc$^{-3}$) can be fitted quite well by fitting
functions of the form
\be
\log \dot{n}_{\rm ion}^{\rm stellar}(z) = 59.786~{\rm e}^{-0.008 z} - 9.844~{\rm e}^{-0.079 z}
\e
which, when combined with the QSO emissivities given in Section~\ref{sec:qso_emissivity}, should be useful
in computing the reionization history.

\section{Thermal evolution of IGM}
\label{sec:temperature}

\begin{figure*}
\centering
  \includegraphics[height=0.45\textwidth, angle=0]{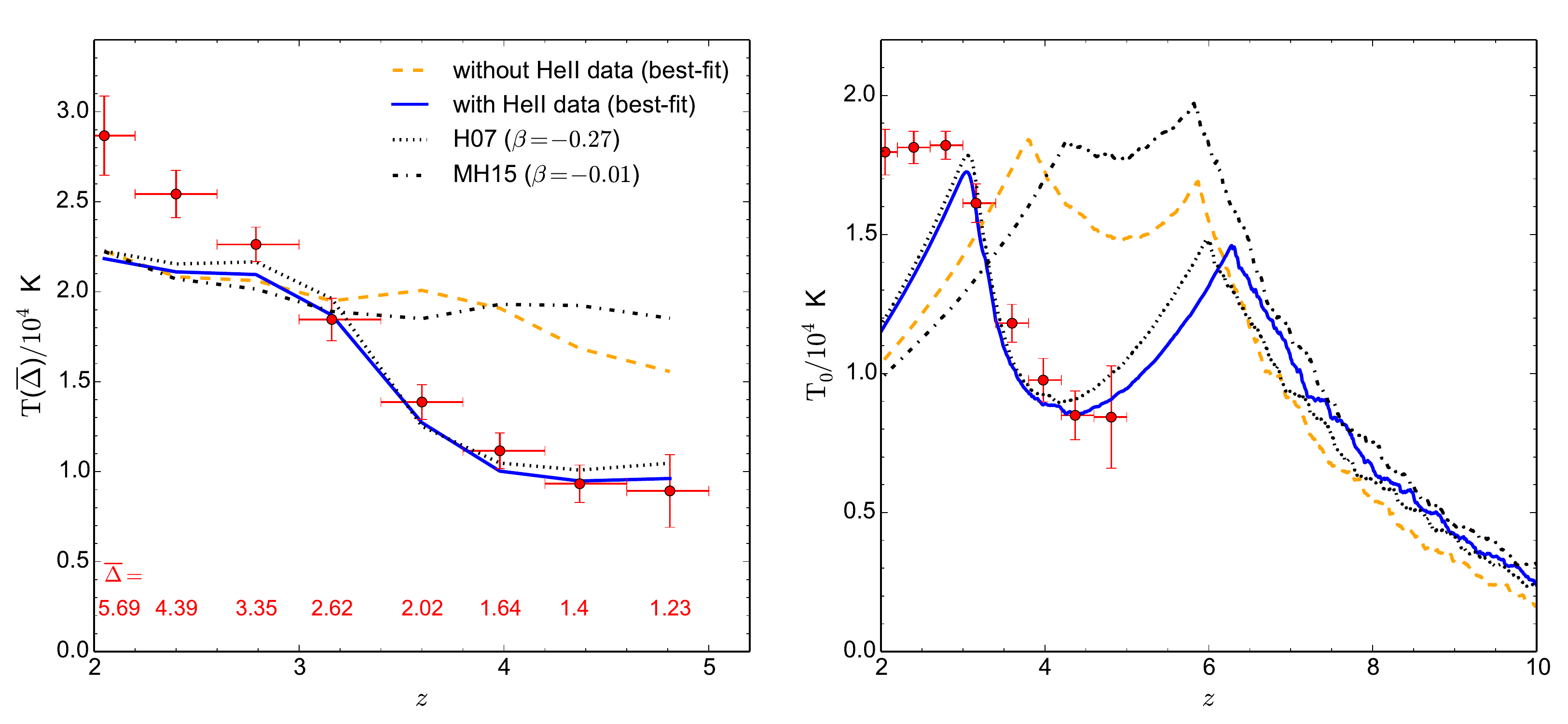}
  \caption{Thermal history for four different reionization models with different
  AGN emissivities; model with $\beta=-0.06$ (orange dashed), $\beta=-0.43$ (blue solid),
  $\beta=-0.27$ (black dotted) and $\beta=-0.01$ (black dot-dashed line).
  {\it Left panel}: evolution of IGM temperature at different overdensities printed
  at the bottom of the plot. The data points shown here are from the Ly$\alpha$ forest
  temperature measurements by \citet{2011MNRAS.410.1096B}.
  {\it Right panel}: temperature of gas at mean IGM density ($\Delta=1$). Data points are 
  inferred from the same temperature measurements, but corrected for the appropriate
  $\gamma$ (slope of the temperature-density relation) of our best-fit model with
  \HeII\ data (solid blue lines).}
\label{fig:temp}
\end{figure*}

A different way of constraining the quasar emissivity and the associated \HeII\ reionization is
by studying the thermal history of the IGM. The AGN-dominated models can lead to an additional
heating of the IGM as they produce significantly more hard ionizing photons with energies in
excess of 4 Ry compared to stellar sources. Thus the observational measurements on the IGM
temperature can also be used to discriminate between different source models and put additional
constraints on reionization \citep{2001ApJ...557..519Z,2008ApJ...689L..81T,
2009ApJ...701...94F,2009ApJ...706L.164C,2012MNRAS.421.1969R,2014MNRAS.443.3761P}. 

Note that a careful calculation of the thermal evolution of the IGM is computationally much more
expensive than that for the semi-analytic models described in the earlier sections and hence is
not suitable for including in the MCMC analysis. Instead, we calculate the thermal histories
for our best-fit models (along with the \citetalias{2007ApJ...654..731H} and \citetalias{2015ApJ...813L...8M} cases)
and compare our predictions with the observations from \cite{2011MNRAS.410.1096B}.

Our model computes the IGM temperature ($T$) evolution self-consistently and separately
for each of the three regions, i.e., (i) completely neutral, (ii) regions where hydrogen is ionized
and helium is singly ionized and (iii) where both species are fully ionized. For neutral regions,
the temperature is usually assumed to decrease adiabatically while in the ionized regions,
it is calculated as \citep{tirth05,2016MNRAS.460.1885U}:
\begin{eqnarray}
 \frac{{\rm d}T}{{\rm d}t} &=&
 -2H(t)T+\frac{2}{3}\frac{T}{\Delta}\frac{{\rm d}\Delta}{{\rm d}t}\nonumber\\ &&
 - \frac{T}{\sum_i X_i}\frac{{\rm d}}{{\rm d}t}\sum_i X_i+
 \frac{2}{3k_Bn_b(1+z)^3}\frac{{\rm d}E}{{\rm d}t}
\end{eqnarray}
where we have defined
\be
 X_i\equiv\frac{n_im_p}{\overline{\rho}_b(\Delta)}
\e
for three independent species $X_{\rm HI}$, $X_{\rm HeI}$, $X_{\rm HeIII}$ and $\overline{\rho}_b$
is the proper mass density of baryons. The first term on the right-hand side accounts for the adiabatic cooling of 
gas due to cosmic expansion. The second term calculates the adiabatic evolution due to collapsing overdensities.
The third term describes the rate of change in the number of particles of the system and in the last term
on the right-hand side, ${\rm d}E/{\rm d}t$ represents the net heating rate per baryon which encodes all
other possible heating and cooling processes. For most purposes, it is sufficient to take into
account only the photoheating, recombination cooling and Compton cooling off CMB photons \citep{tirth05}.

In order to calculate the temperature at a given density ${\rm \Delta}$
and redshift $z$, we track the thermal evolution of a large number of density elements ($\sim50, 000$),
generated according to the IGM probability distribution. It is this step which increases the computational
cost of the analysis. During hydrogen reionization, our model assumes that all regions with
$\Delta > \Delta_{\rm crit}$ remain neutral, while a fraction $Q_{\rm HII}$ of $\Delta < \Delta_{\rm crit}$
regions are ionized \citep{2000ApJ...530....1M}. In the post-reionization $Q_{\rm HII} = 1$ era,
the effect of ionizing radiation is to increase the value of the density threshold above which
the IGM is neutral. At every time step, we calculate the increase in the ionized fraction and randomly
assign the corresponding number of density elements as newly ionized. Following \cite{2016MNRAS.460.1885U}
and \cite{2017MNRAS.468.4691D} (and the references therein), each of these density elements is
instantaneously heated, on top of the uniform photoheating background, to a temperature of $20,000$ K.
We follow an identical procedure for \HeII\ reionization as well and assume the density elements to be
additionally heated by $8,000$ K when they reach their \HeII\ reionization redshifts. 

This method allows us to calculate, at any given redshift, the dependence of the gas temperature
$T(\Delta)$ on the overdensity $\Delta$. We can fit the $T(\Delta)$ relation using a power-law of the form
\be
 T(\Delta)=T_0~\Delta^{\gamma-1},
\e
where $T_0$ is the temperature of the mean density ($\Delta = 1$) gas and 
$\gamma$ is the slope of the temperature-density relation. We can estimate the values of $T_0$ and
$\gamma$ at every redshift.

In the {\it left} panel of Fig.~\ref{fig:temp}, we show the redshift evolution of IGM temperature
$T(\overline{\Delta})$ at different overdensities $\overline{\Delta}$ at redshifts where the data
points from \cite{2011MNRAS.410.1096B} exist. Different curves are for different quasar emissivity
models, while the points with error bars are the observed data \citep{2011MNRAS.410.1096B}.
The first point to note is that all the models, irrespective of their quasar emissivity, underpredict
the temperature at $z < 3$. This could be because of assumptions in our semi-analytic model
(e.g., the effective spectral index of the radiation from the quasars could be harder than what we assumed,
which would lead to additional heating), or because of additional heating sources like the blazars
\citep{2012MNRAS.423..149P}, cosmic rays or intergalactic dust absorption \citep{2016MNRAS.460.1885U}.
Given the fact that our calculations of the temperature do not match the observations at $z < 3$,
any conclusions drawn in this section should be interpreted with caution.

We find that the \citetalias{2007ApJ...654..731H} model and our best-fit model with \HeII\ data
are in excellent agreement with the observations at $z\gtrsim3$. On the other hand, models with a
more quasar contribution, i.e., the \citetalias{2015ApJ...813L...8M} and the best-fit model without
\HeII\ data, predict considerably higher temperature than what is observed at these redshifts.
These models have considerably earlier \HeII\ reionization and hence the IGM is significantly
photoheated at $3 \lesssim z \lesssim 5$.

The same trends can be seen from the evolution of temperature at the mean density $T_0(z)$
({\it right} panel). The two peaks seen in all the models correspond to the hydrogen and \HeII\ reionization
respectively. Also shown are the data points from the same observation by \cite{2011MNRAS.410.1096B},
but adjusted accordingly using the temperature-density relation of the best-fit model with \HeII\ data
(solid blue lines). Models with higher $\beta$ heat up the IGM at much earlier epoch than the what the
observations show. Our main findings are in excellent agreement with the recent study by \cite{2017MNRAS.468.4691D},
where they conclude that the AGN-dominated scenario struggles to match the temperature measurements.

\section{Conclusions}
\label{sec:conclusions}

The new Planck results along with the recent discovery of significantly high number of faint
AGNs from multiwavelength deep surveys at higher redshifts \citep{2012A&A...537A..16F,2015A&A...578A..83G}
demand careful investigation of the actual contribution from star-forming galaxies and quasars as
reionization sources. In addition, various current studies suggest that the globally-averaged
escape fraction of ionizing photons is very small $\sim$ a few percent \citep{2014ARA&A..52..415M,2015MNRAS.453..960M,
mitra4,2016MNRAS.457.1550H}. Based on this idea, a QSO dominated reionization scenario has been
explored in many contemporary studies \citep{2015ApJ...813L...8M,2016MNRAS.457.4051K,2016arXiv160204407Y}.
On the other hand, \cite{2015ApJ...813L..35K} and \cite{2016ApJ...833..222J}
claimed that the number of spectroscopically
identified faint AGNs may not be high enough to fully account for the reionization of the IGM.
Recently, \cite{2017MNRAS.468.4691D} also indicates a possible drawback of such
high AGN emissivity models. This makes the claim of quasar-dominated scenario debatable.
Thus, it would be interesting to see how our data-constrained semi-analytical reionization model
can distinguish between the models with different comoving AGN emissivities.
In this work, we have expanded our previous model \citep{mitra4} and evaluated the impact of
QSOs on reionization for different cases. As the observations on AGN emissivities at $z>3$
are still very uncertain, we model the AGN contribution in a way that its evolution at
high redshifts ($z>2$) will solely be determined by a single parameter $\beta$.
We check that, $\beta=-0.27$ and $\beta=-0.01$ can reasonably restore our old model with the
observed quasar luminosity functions from \cite{2007ApJ...654..731H} and a quasar-dominated model with
best-fit AGN emissivity from \cite{2015ApJ...813L...8M} respectively. We denote the former case
as \citetalias{2007ApJ...654..731H} and the later one as \citetalias{2015ApJ...813L...8M}.
We try to constrain the value of $\beta$ against different observations related to hydrogen and helium
reionization: measurements of effective \HI\ Ly$\alpha$ optical depth $\tau_{\rm eff, HI}$, redshift evolution of LLS
$\de N_{\rm LL}/\de z$, reionization optical depth from Planck 2016 and \HeII\ Ly$\alpha$
forest data.

\begin{itemize}

 \item We find that if we do not include \HeII\ data in our analysis, a wide
 range of $\beta$-model can be allowed by the current observations. Both the quasar-dominated
 ($\beta\sim0$) and stellar-dominated ($\beta\sim-4$; negligible QSOs at $z>2$) scenarios are possible
 within the 2-$\sigma$ confidence limits. Any variation in the quasar emissivity is appropriately
 compensated by the radiation from the stellar sources, thus ensuring that the constraints arising
 from hydrogen reionization data are not violated. It is thus not possible to put any meaningful
 constraints on the quasar emissivity using hydrogen reionization alone.

\item In contrast, using \HeII\ data in the analysis
 can put a tighter constraint on $\beta$ and rules out the possibility of quasar dominated models
 of hydrogen reionization. This model supports the decade-old {\it two-component} view of
 cosmic reionization; quasars contribute significantly at lower redshifts and later the star-forming
 early galaxies take over. An escape fraction of $8$ to $13\%$ from those galaxies at
 $z\gtrsim6$ are needed to provide the appropriate ionizing flux. The fraction of photoionization
 rate contributed by the quasars cannot be larger that $\sim 10 \% (\sim 50 \%)$ at $z = 6 (4)$.
 
 \item The models where AGNs throughout dominate the reionization process
 (e.g. \citetalias{2015ApJ...813L...8M} model) favour an early helium reionization
 (completed at $z\gtrsim4$) history. They produce a much more slowly evolving
 \HeII\ effective  optical depth than observation \citep{2016ApJ...825..144W}
 suggests. Also such models are inconsistent with the
 observational measurements of IGM temperature \citep{2011MNRAS.410.1096B}.
 They heat up the IGM at much earlier epoch than the actual observation indicates.

 \item On the other hand, the model constrained by \HeII\ data has no difficulties in
 matching both the $\tau_{\rm eff, HeII}$ and temperature data. The best-fit model
 predicts a relatively late helium reionization ending at $z\approx3$. $\tau_{\rm eff, HeII}$
 increases rapidly from $z=2.4$ to $z=3.4$ by a factor of $\sim6$. Our old
 \citetalias{2007ApJ...654..731H} case is well within the 2-$\sigma$ limits
 of such models.

\end{itemize}

As a final note, we should mention some caveats of our model.
The star-formation efficiency and escape fraction parameters should
depend on the halo mass and/or redshifts (however weakly). For simplicity, we take them
as a constant throughout this work. A detailed probe of parameter space using
the principal component analysis, at the expense of larger computational time, might
be useful in this case. We notice that, our simplified $\lambda_{\rm mfp}$ prescription
starts to break down when it becomes comparable to the Hubble radius at low-redshift regime
($z\lesssim2$). Perhaps a more accurate mean free path calculation will be needful. 
In our model, following \cite{2000ApJ...530....1M},
we assume that reionization is said to be complete once all the low-density regions
with $\Delta<\Delta_{\rm crit}$ are ionized, while the higher density gas is still neutral.
This postulate might not be fully correct, as in a real scenario, the ionization factor can
depend on the local intensity of ionizing photons and/or the degree of self-shielding
\citep{2000ApJ...530....1M,tirth09}.
Furthermore, a proper account of the quasar proximity effect (where the ionization rate
around a quasar exceeds the background flux) in our model might provide additional
insights \citep{1988ApJ...327..570B,1994MNRAS.266..343M,2014MNRAS.443.3761P,2017MNRAS.468.4691D}.
Finally, recent observations \citep{2015MNRAS.447.3402B} indicate a rapid dispersion
in Ly-$\alpha$ forest effective optical depth at $z>5$. \cite{2016arXiv161102711D} argued that
matching such large scatter requires a much shorter (factor of $\gtrsim3$) spatially averaged
mean free path than the actual measured value from \cite{2014MNRAS.445.1745W} at high redshifts,
assuming galaxies to be the dominant sources of ionizing photons \citep{2016MNRAS.460.1328D}.
However, they also claimed that these opacity fluctuations might be driven by the residual
inhomogeneities in the IGM temperature arising from a patchy reionization process rather than by
the ionization background itself. A more rigorous modeling of IGM along with a careful estimation
of mean free path is needed in our analysis to fully include these effects.
Also, our model for temperature evolution is probably not very
accurate at low redshifts ($z<3$). Additional sources of heating like blazars might also be
required, which we have not considered in this work.

Nonetheless, future discovery of faint quasars at high redshifts might be worthwhile to
clearly distinguish between the models with different AGN contribution. A refined
reionization model along with the improved measurements on intergalactic \HeII\ Ly$\alpha$
absorption spectra and high-redshift temperature measurements will be needed to put a more
decisive constraint on the helium reionization scenario.

\section*{Acknowledgements}
We thank the anonymous referee for constructive suggestions that improved this paper.
We would also like to thank Vikram Khaire for useful discussions.

\bibliography{reionization-smitra}

\begin{thebibliography}{}
\makeatletter
\relax
\def\mn@urlcharsother{\let\do\@makeother \do\$\do\&\do\#\do\^\do\_\do\%\do\~}
\def\mn@doi{\begingroup\mn@urlcharsother \@ifnextchar [ {\mn@doi@}
  {\mn@doi@[]}}
\def\mn@doi@[#1]#2{\def\@tempa{#1}\ifx\@tempa\@empty \href
  {http://dx.doi.org/#2} {doi:#2}\else \href {http://dx.doi.org/#2} {#1}\fi
  \endgroup}
\def\mn@eprint#1#2{\mn@eprint@#1:#2::\@nil}
\def\mn@eprint@arXiv#1{\href {http://arxiv.org/abs/#1} {{\tt arXiv:#1}}}
\def\mn@eprint@dblp#1{\href {http://dblp.uni-trier.de/rec/bibtex/#1.xml}
  {dblp:#1}}
\def\mn@eprint@#1:#2:#3:#4\@nil{\def\@tempa {#1}\def\@tempb {#2}\def\@tempc
  {#3}\ifx \@tempc \@empty \let \@tempc \@tempb \let \@tempb \@tempa \fi \ifx
  \@tempb \@empty \def\@tempb {arXiv}\fi \@ifundefined
  {mn@eprint@\@tempb}{\@tempb:\@tempc}{\expandafter \expandafter \csname
  mn@eprint@\@tempb\endcsname \expandafter{\@tempc}}}

\bibitem[\protect\citeauthoryear{{Bajtlik}, {Duncan}  \& {Ostriker}}{{Bajtlik}
  et~al.}{1988}]{1988ApJ...327..570B}
{Bajtlik} S.,  {Duncan} R.~C.,   {Ostriker} J.~P.,  1988, \mn@doi [\apj]
  {10.1086/166217}, \href {http://adsabs.harvard.edu/abs/1988ApJ...327..570B}
  {327, 570}

\bibitem[\protect\citeauthoryear{{Barkana} \& {Loeb}}{{Barkana} \&
  {Loeb}}{2001}]{BarkanaLoeb01}
{Barkana} R.,  {Loeb} A.,  2001, \mn@doi [\physrep]
  {10.1016/S0370-1573(01)00019-9}, \href
  {http://adsabs.harvard.edu/abs/2001PhR...349..125B} {349, 125}

\bibitem[\protect\citeauthoryear{{Becker} \& {Bolton}}{{Becker} \&
  {Bolton}}{2013}]{2013MNRAS.436.1023B}
{Becker} G.~D.,  {Bolton} J.~S.,  2013, \mn@doi [\mnras]
  {10.1093/mnras/stt1610}, \href
  {http://adsabs.harvard.edu/abs/2013MNRAS.436.1023B} {436, 1023}

\bibitem[\protect\citeauthoryear{Becker et~al.}{Becker
  et~al.}{2001}]{Becker:2001ee}
Becker R.~H.,  et~al., 2001, \mn@doi [Astron.J.] {10.1086/324231}, 122, 2850

\bibitem[\protect\citeauthoryear{{Becker}, {Bolton}, {Haehnelt}  \&
  {Sargent}}{{Becker} et~al.}{2011}]{2011MNRAS.410.1096B}
{Becker} G.~D.,  {Bolton} J.~S.,  {Haehnelt} M.~G.,   {Sargent} W.~L.~W.,
  2011, \mn@doi [\mnras] {10.1111/j.1365-2966.2010.17507.x}, \href
  {http://adsabs.harvard.edu/abs/2011MNRAS.410.1096B} {410, 1096}

\bibitem[\protect\citeauthoryear{{Becker}, {Hewett}, {Worseck}  \&
  {Prochaska}}{{Becker} et~al.}{2013}]{2013MNRAS.430.2067B}
{Becker} G.~D.,  {Hewett} P.~C.,  {Worseck} G.,   {Prochaska} J.~X.,  2013,
  \mn@doi [\mnras] {10.1093/mnras/stt031}, \href
  {http://adsabs.harvard.edu/abs/2013MNRAS.430.2067B} {430, 2067}

\bibitem[\protect\citeauthoryear{{Becker}, {Bolton}, {Madau}, {Pettini},
  {Ryan-Weber}  \& {Venemans}}{{Becker} et~al.}{2015}]{2015MNRAS.447.3402B}
{Becker} G.~D.,  {Bolton} J.~S.,  {Madau} P.,  {Pettini} M.,  {Ryan-Weber}
  E.~V.,   {Venemans} B.~P.,  2015, \mn@doi [\mnras] {10.1093/mnras/stu2646},
  \href {http://adsabs.harvard.edu/abs/2015MNRAS.447.3402B} {447, 3402}

\bibitem[\protect\citeauthoryear{{Bennett} et~al.}{{Bennett}
  et~al.}{2013}]{2013ApJS..208...20B}
{Bennett} C.~L.,  et~al., 2013, \mn@doi [\apjs] {10.1088/0067-0049/208/2/20},
  \href {http://adsabs.harvard.edu/abs/2013ApJS..208...20B} {208, 20}

\bibitem[\protect\citeauthoryear{{Bolton} \& {Haehnelt}}{{Bolton} \&
  {Haehnelt}}{2007}]{2007MNRAS.382..325B}
{Bolton} J.~S.,  {Haehnelt} M.~G.,  2007, \mn@doi [\mnras]
  {10.1111/j.1365-2966.2007.12372.x}, \href
  {http://adsabs.harvard.edu/abs/2007MNRAS.382..325B} {382, 325}

\bibitem[\protect\citeauthoryear{{Bolton} et~al.}{{Bolton}
  et~al.}{2011}]{2011MNRAS.416L..70B}
{Bolton} J.~S.,  et~al., 2011, \mn@doi [\mnras]
  {10.1111/j.1745-3933.2011.01100.x}, \href
  {http://adsabs.harvard.edu/abs/2011MNRAS.416L..70B} {416, L70}

\bibitem[\protect\citeauthoryear{{Bouwens} \& {Illingworth}}{{Bouwens} \&
  {Illingworth}}{2006}]{2006NewAR..50..152B}
{Bouwens} R.,  {Illingworth} G.,  2006, \mn@doi [\nar]
  {10.1016/j.newar.2005.11.027}, \href
  {http://adsabs.harvard.edu/abs/2006NewAR..50..152B} {50, 152}

\bibitem[\protect\citeauthoryear{{Bouwens} et~al.}{{Bouwens}
  et~al.}{2015a}]{2015ApJ...803...34B}
{Bouwens} R.~J.,  et~al., 2015a, \mn@doi [\apj] {10.1088/0004-637X/803/1/34},
  \href {http://adsabs.harvard.edu/abs/2015ApJ...803...34B} {803, 34}

\bibitem[\protect\citeauthoryear{{Bouwens}, {Illingworth}, {Oesch}, {Caruana},
  {Holwerda}, {Smit}  \& {Wilkins}}{{Bouwens}
  et~al.}{2015b}]{2015ApJ...811..140B}
{Bouwens} R.~J.,  {Illingworth} G.~D.,  {Oesch} P.~A.,  {Caruana} J.,
  {Holwerda} B.,  {Smit} R.,   {Wilkins} S.,  2015b, \mn@doi [\apj]
  {10.1088/0004-637X/811/2/140}, \href
  {http://adsabs.harvard.edu/abs/2015ApJ...811..140B} {811, 140}

\bibitem[\protect\citeauthoryear{{Bowler} et~al.}{{Bowler}
  et~al.}{2014}]{2014arXiv1411.2976B}
{Bowler} R.~A.~A.,  et~al., 2014, arXiv:1411.2976, \href
  {http://adsabs.harvard.edu/abs/2014arXiv1411.2976B} {}

\bibitem[\protect\citeauthoryear{{Bradley} et~al.}{{Bradley}
  et~al.}{2012}]{2012ApJ...760..108B}
{Bradley} L.~D.,  et~al., 2012, \mn@doi [\apj] {10.1088/0004-637X/760/2/108},
  \href {http://adsabs.harvard.edu/abs/2012ApJ...760..108B} {760, 108}

\bibitem[\protect\citeauthoryear{{Cai}, {Lapi}, {Bressan}, {De Zotti},
  {Negrello}  \& {Danese}}{{Cai} et~al.}{2014}]{2014ApJ...785...65C}
{Cai} Z.-Y.,  {Lapi} A.,  {Bressan} A.,  {De Zotti} G.,  {Negrello} M.,
  {Danese} L.,  2014, \mn@doi [\apj] {10.1088/0004-637X/785/1/65}, \href
  {http://adsabs.harvard.edu/abs/2014ApJ...785...65C} {785, 65}

\bibitem[\protect\citeauthoryear{{Cen}, {McDonald}, {Trac}  \& {Loeb}}{{Cen}
  et~al.}{2009}]{2009ApJ...706L.164C}
{Cen} R.,  {McDonald} P.,  {Trac} H.,   {Loeb} A.,  2009, \mn@doi [\apjl]
  {10.1088/0004-637X/706/1/L164}, \href
  {http://adsabs.harvard.edu/abs/2009ApJ...706L.164C} {706, L164}

\bibitem[\protect\citeauthoryear{{Chornock} et~al.}{{Chornock}
  et~al.}{2013}]{2013ApJ...774...26C}
{Chornock} R.,  et~al., 2013, \mn@doi [\apj] {10.1088/0004-637X/774/1/26},
  \href {http://adsabs.harvard.edu/abs/2013ApJ...774...26C} {774, 26}

\bibitem[\protect\citeauthoryear{{Choudhury}}{{Choudhury}}{2009}]{tirth09}
{Choudhury} T.~R.,  2009, Current Science, \href
  {http://adsabs.harvard.edu/abs/2009CSci...97..841C} {97, 841}

\bibitem[\protect\citeauthoryear{{Choudhury} \& {Ferrara}}{{Choudhury} \&
  {Ferrara}}{2005}]{tirth05}
{Choudhury} T.~R.,  {Ferrara} A.,  2005, \mn@doi [\mnras]
  {10.1111/j.1365-2966.2005.09196.x}, \href
  {http://adsabs.harvard.edu/abs/2005MNRAS.361..577C} {361, 577}

\bibitem[\protect\citeauthoryear{{Choudhury} \& {Ferrara}}{{Choudhury} \&
  {Ferrara}}{2006a}]{tirth06a}
{Choudhury} T.~R.,  {Ferrara} A.,  2006a, preprint, \href
  {http://adsabs.harvard.edu/abs/2006astro.ph..3149C} {} (\mn@eprint {arXiv}
  {arXiv:astro-ph/0603149})

\bibitem[\protect\citeauthoryear{{Choudhury} \& {Ferrara}}{{Choudhury} \&
  {Ferrara}}{2006b}]{tirth06}
{Choudhury} T.~R.,  {Ferrara} A.,  2006b, \mn@doi [\mnras]
  {10.1111/j.1745-3933.2006.00207.x}, \href
  {http://adsabs.harvard.edu/abs/2006MNRAS.371L..55C} {371, L55}

\bibitem[\protect\citeauthoryear{{Choudhury}, {Puchwein}, {Haehnelt}  \&
  {Bolton}}{{Choudhury} et~al.}{2014}]{2014arXiv1412.4790C}
{Choudhury} T.~R.,  {Puchwein} E.,  {Haehnelt} M.~G.,   {Bolton} J.~S.,  2014,
  arXiv:1412.4790, \href {http://adsabs.harvard.edu/abs/2014arXiv1412.4790C} {}

\bibitem[\protect\citeauthoryear{{Civano} et~al.,}{{Civano}
  et~al.}{2011}]{2011ApJ...741...91C}
{Civano} F.,  et~al., 2011, \mn@doi [\apj] {10.1088/0004-637X/741/2/91}, \href
  {http://adsabs.harvard.edu/abs/2011ApJ...741...91C} {741, 91}

\bibitem[\protect\citeauthoryear{{Compostella}, {Cantalupo}  \&
  {Porciani}}{{Compostella} et~al.}{2013}]{2013MNRAS.435.3169C}
{Compostella} M.,  {Cantalupo} S.,   {Porciani} C.,  2013, \mn@doi [\mnras]
  {10.1093/mnras/stt1510}, \href
  {http://adsabs.harvard.edu/abs/2013MNRAS.435.3169C} {435, 3169}

\bibitem[\protect\citeauthoryear{{Compostella}, {Cantalupo}  \&
  {Porciani}}{{Compostella} et~al.}{2014}]{2014MNRAS.445.4186C}
{Compostella} M.,  {Cantalupo} S.,   {Porciani} C.,  2014, \mn@doi [\mnras]
  {10.1093/mnras/stu2035}, \href
  {http://adsabs.harvard.edu/abs/2014MNRAS.445.4186C} {445, 4186}

\bibitem[\protect\citeauthoryear{{D'Aloisio}, {McQuinn}, {Davies}  \&
  {Furlanetto}}{{D'Aloisio} et~al.}{2016}]{2016arXiv161102711D}
{D'Aloisio} A.,  {McQuinn} M.,  {Davies} F.~B.,   {Furlanetto} S.~R.,  2016,
  preprint, \href {http://adsabs.harvard.edu/abs/2016arXiv161102711D} {}
  (\mn@eprint {arXiv} {1611.02711})

\bibitem[\protect\citeauthoryear{{D'Aloisio}, {Upton Sanderbeck}, {McQuinn},
  {Trac}  \& {Shapiro}}{{D'Aloisio} et~al.}{2017}]{2017MNRAS.468.4691D}
{D'Aloisio} A.,  {Upton Sanderbeck} P.~R.,  {McQuinn} M.,  {Trac} H.,
  {Shapiro} P.~R.,  2017, \mn@doi [\mnras] {10.1093/mnras/stx711}, \href
  {http://adsabs.harvard.edu/abs/2017MNRAS.468.4691D} {468, 4691}

\bibitem[\protect\citeauthoryear{{Davies} \& {Furlanetto}}{{Davies} \&
  {Furlanetto}}{2016}]{2016MNRAS.460.1328D}
{Davies} F.~B.,  {Furlanetto} S.~R.,  2016, \mn@doi [\mnras]
  {10.1093/mnras/stw931}, \href
  {http://adsabs.harvard.edu/abs/2016MNRAS.460.1328D} {460, 1328}

\bibitem[\protect\citeauthoryear{{Dijkstra}, {Wyithe}  \& {Haiman}}{{Dijkstra}
  et~al.}{2007}]{2007MNRAS.379..253D}
{Dijkstra} M.,  {Wyithe} J.~S.~B.,   {Haiman} Z.,  2007, \mn@doi [\mnras]
  {10.1111/j.1365-2966.2007.11936.x}, \href
  {http://adsabs.harvard.edu/abs/2007MNRAS.379..253D} {379, 253}

\bibitem[\protect\citeauthoryear{{Dixon} \& {Furlanetto}}{{Dixon} \&
  {Furlanetto}}{2009}]{2009ApJ...706..970D}
{Dixon} K.~L.,  {Furlanetto} S.~R.,  2009, \mn@doi [\apj]
  {10.1088/0004-637X/706/2/970}, \href
  {http://adsabs.harvard.edu/abs/2009ApJ...706..970D} {706, 970}

\bibitem[\protect\citeauthoryear{{Fan} et~al.}{{Fan}
  et~al.}{2006}]{2006AJ....132..117F}
{Fan} X.,  et~al., 2006, \mn@doi [\aj] {10.1086/504836}, \href
  {http://adsabs.harvard.edu/abs/2006AJ....132..117F} {132, 117}

\bibitem[\protect\citeauthoryear{{Faucher-Gigu{\`e}re}, {Lidz}, {Zaldarriaga}
  \& {Hernquist}}{{Faucher-Gigu{\`e}re} et~al.}{2009}]{2009ApJ...703.1416F}
{Faucher-Gigu{\`e}re} C.-A.,  {Lidz} A.,  {Zaldarriaga} M.,   {Hernquist} L.,
  2009, \mn@doi [\apj] {10.1088/0004-637X/703/2/1416}, \href
  {http://adsabs.harvard.edu/abs/2009ApJ...703.1416F} {703, 1416}

\bibitem[\protect\citeauthoryear{{Finlator}, {Oppenheimer}, {Dav{\'e}},
  {Zackrisson}, {Thompson}  \& {Huang}}{{Finlator}
  et~al.}{2016}]{2016MNRAS.459.2299F}
{Finlator} K.,  {Oppenheimer} B.~D.,  {Dav{\'e}} R.,  {Zackrisson} E.,
  {Thompson} R.,   {Huang} S.,  2016, \mn@doi [\mnras] {10.1093/mnras/stw805},
  \href {http://adsabs.harvard.edu/abs/2016MNRAS.459.2299F} {459, 2299}

\bibitem[\protect\citeauthoryear{{Fiore} et~al.,}{{Fiore}
  et~al.}{2012}]{2012A&A...537A..16F}
{Fiore} F.,  et~al., 2012, \mn@doi [\aap] {10.1051/0004-6361/201117581}, \href
  {http://adsabs.harvard.edu/abs/2012A%26A...537A..16F} {537, A16}

\bibitem[\protect\citeauthoryear{{Furlanetto} \& {Oh}}{{Furlanetto} \&
  {Oh}}{2009}]{2009ApJ...701...94F}
{Furlanetto} S.~R.,  {Oh} S.~P.,  2009, \mn@doi [\apj]
  {10.1088/0004-637X/701/1/94}, \href
  {http://adsabs.harvard.edu/abs/2009ApJ...701...94F} {701, 94}

\bibitem[\protect\citeauthoryear{{Gallerani}, {Choudhury}  \&
  {Ferrara}}{{Gallerani} et~al.}{2006}]{2006MNRAS.370.1401G}
{Gallerani} S.,  {Choudhury} T.~R.,   {Ferrara} A.,  2006, \mn@doi [\mnras]
  {10.1111/j.1365-2966.2006.10553.x}, \href
  {http://adsabs.harvard.edu/abs/2006MNRAS.370.1401G} {370, 1401}

\bibitem[\protect\citeauthoryear{{Giallongo}, {Menci}, {Fiore}, {Castellano},
  {Fontana}, {Grazian}  \& {Pentericci}}{{Giallongo}
  et~al.}{2012}]{2012ApJ...755..124G}
{Giallongo} E.,  {Menci} N.,  {Fiore} F.,  {Castellano} M.,  {Fontana} A.,
  {Grazian} A.,   {Pentericci} L.,  2012, \mn@doi [\apj]
  {10.1088/0004-637X/755/2/124}, \href
  {http://adsabs.harvard.edu/abs/2012ApJ...755..124G} {755, 124}

\bibitem[\protect\citeauthoryear{{Giallongo} et~al.,}{{Giallongo}
  et~al.}{2015}]{2015A&A...578A..83G}
{Giallongo} E.,  et~al., 2015, \mn@doi [\aap] {10.1051/0004-6361/201425334},
  \href {http://adsabs.harvard.edu/abs/2015A%26A...578A..83G} {578, A83}

\bibitem[\protect\citeauthoryear{{Gleser}, {Nusser}, {Benson}, {Ohno}  \&
  {Sugiyama}}{{Gleser} et~al.}{2005}]{2005MNRAS.361.1399G}
{Gleser} L.,  {Nusser} A.,  {Benson} A.~J.,  {Ohno} H.,   {Sugiyama} N.,  2005,
  \mn@doi [\mnras] {10.1111/j.1365-2966.2005.09276.x}, \href
  {http://adsabs.harvard.edu/abs/2005MNRAS.361.1399G} {361, 1399}

\bibitem[\protect\citeauthoryear{{Glikman}, {Djorgovski}, {Stern}, {Dey},
  {Jannuzi}  \& {Lee}}{{Glikman} et~al.}{2011}]{2011ApJ...728L..26G}
{Glikman} E.,  {Djorgovski} S.~G.,  {Stern} D.,  {Dey} A.,  {Jannuzi} B.~T.,
  {Lee} K.-S.,  2011, \mn@doi [\apjl] {10.1088/2041-8205/728/2/L26}, \href
  {http://adsabs.harvard.edu/abs/2011ApJ...728L..26G} {728, L26}

\bibitem[\protect\citeauthoryear{{Hassan}, {Dav{\'e}}, {Finlator}  \&
  {Santos}}{{Hassan} et~al.}{2016}]{2016MNRAS.457.1550H}
{Hassan} S.,  {Dav{\'e}} R.,  {Finlator} K.,   {Santos} M.~G.,  2016, \mn@doi
  [\mnras] {10.1093/mnras/stv3001}, \href
  {http://adsabs.harvard.edu/abs/2016MNRAS.457.1550H} {457, 1550}

\bibitem[\protect\citeauthoryear{{Hopkins}, {Richards}  \&
  {Hernquist}}{{Hopkins} et~al.}{2007}]{2007ApJ...654..731H}
{Hopkins} P.~F.,  {Richards} G.~T.,   {Hernquist} L.,  2007, \mn@doi [\apj]
  {10.1086/509629}, \href {http://adsabs.harvard.edu/abs/2007ApJ...654..731H}
  {654, 731}

\bibitem[\protect\citeauthoryear{{Iliev}, {Shapiro}, {McDonald}, {Mellema}  \&
  {Pen}}{{Iliev} et~al.}{2008}]{2008MNRAS.391...63I}
{Iliev} I.~T.,  {Shapiro} P.~R.,  {McDonald} P.,  {Mellema} G.,   {Pen} U.-L.,
  2008, \mn@doi [\mnras] {10.1111/j.1365-2966.2008.13879.x}, \href
  {http://adsabs.harvard.edu/abs/2008MNRAS.391...63I} {391, 63}

\bibitem[\protect\citeauthoryear{{Jiang} et~al.,}{{Jiang}
  et~al.}{2016}]{2016ApJ...833..222J}
{Jiang} L.,  et~al., 2016, \mn@doi [\apj] {10.3847/1538-4357/833/2/222}, \href
  {http://adsabs.harvard.edu/abs/2016ApJ...833..222J} {833, 222}

\bibitem[\protect\citeauthoryear{{Khaire} \& {Srianand}}{{Khaire} \&
  {Srianand}}{2013}]{2013MNRAS.431L..53K}
{Khaire} V.,  {Srianand} R.,  2013, \mn@doi [\mnras] {10.1093/mnrasl/slt007},
  \href {http://adsabs.harvard.edu/abs/2013MNRAS.431L..53K} {431, L53}

\bibitem[\protect\citeauthoryear{{Khaire}, {Srianand}, {Choudhury}  \&
  {Gaikwad}}{{Khaire} et~al.}{2016}]{2016MNRAS.457.4051K}
{Khaire} V.,  {Srianand} R.,  {Choudhury} T.~R.,   {Gaikwad} P.,  2016, \mn@doi
  [\mnras] {10.1093/mnras/stw192}, \href
  {http://adsabs.harvard.edu/abs/2016MNRAS.457.4051K} {457, 4051}

\bibitem[\protect\citeauthoryear{{Kim} et~al.,}{{Kim}
  et~al.}{2015}]{2015ApJ...813L..35K}
{Kim} Y.,  et~al., 2015, \mn@doi [\apjl] {10.1088/2041-8205/813/2/L35}, \href
  {http://adsabs.harvard.edu/abs/2015ApJ...813L..35K} {813, L35}

\bibitem[\protect\citeauthoryear{Komatsu et~al.}{Komatsu
  et~al.}{2011}]{Komatsu:2010fb}
Komatsu E.,  et~al., 2011, \mn@doi [Astrophys.J.Suppl.]
  {10.1088/0067-0049/192/2/18}, 192, 18

\bibitem[\protect\citeauthoryear{{Kulkarni} \& {Choudhury}}{{Kulkarni} \&
  {Choudhury}}{2011}]{2011MNRAS.412.2781K}
{Kulkarni} G.,  {Choudhury} T.~R.,  2011, \mn@doi [\mnras]
  {10.1111/j.1365-2966.2010.18100.x}, \href
  {http://adsabs.harvard.edu/abs/2011MNRAS.412.2781K} {412, 2781}

\bibitem[\protect\citeauthoryear{{Lewis} \& {Bridle}}{{Lewis} \&
  {Bridle}}{2002}]{2002PhRvD..66j3511L}
{Lewis} A.,  {Bridle} S.,  2002, \mn@doi [\prd] {10.1103/PhysRevD.66.103511},
  \href {http://adsabs.harvard.edu/abs/2002PhRvD..66j3511L} {66, 103511}

\bibitem[\protect\citeauthoryear{{Loeb}}{{Loeb}}{2006}]{2006astro.ph..3360L}
{Loeb} A.,  2006, preprint, \href
  {http://adsabs.harvard.edu/abs/2006astro.ph..3360L} {} (\mn@eprint {arXiv}
  {arXiv:astro-ph/0603360})

\bibitem[\protect\citeauthoryear{{Loeb} \& {Barkana}}{{Loeb} \&
  {Barkana}}{2001}]{LoebBarkana01}
{Loeb} A.,  {Barkana} R.,  2001, \mn@doi [\araa]
  {10.1146/annurev.astro.39.1.19}, \href
  {http://adsabs.harvard.edu/abs/2001ARA%26A..39...19L} {39, 19}

\bibitem[\protect\citeauthoryear{{Ma}, {Kasen}, {Hopkins},
  {Faucher-Gigu{\`e}re}, {Quataert}, {Kere{\v s}}  \& {Murray}}{{Ma}
  et~al.}{2015}]{2015MNRAS.453..960M}
{Ma} X.,  {Kasen} D.,  {Hopkins} P.~F.,  {Faucher-Gigu{\`e}re} C.-A.,
  {Quataert} E.,  {Kere{\v s}} D.,   {Murray} N.,  2015, \mn@doi [\mnras]
  {10.1093/mnras/stv1679}, \href
  {http://adsabs.harvard.edu/abs/2015MNRAS.453..960M} {453, 960}

\bibitem[\protect\citeauthoryear{{Madau} \& {Dickinson}}{{Madau} \&
  {Dickinson}}{2014}]{2014ARA&A..52..415M}
{Madau} P.,  {Dickinson} M.,  2014, \mn@doi [\araa]
  {10.1146/annurev-astro-081811-125615}, \href
  {http://adsabs.harvard.edu/abs/2014ARA%26A..52..415M} {52, 415}

\bibitem[\protect\citeauthoryear{{Madau} \& {Haardt}}{{Madau} \&
  {Haardt}}{2015}]{2015ApJ...813L...8M}
{Madau} P.,  {Haardt} F.,  2015, \mn@doi [\apjl] {10.1088/2041-8205/813/1/L8},
  \href {http://adsabs.harvard.edu/abs/2015ApJ...813L...8M} {813, L8}

\bibitem[\protect\citeauthoryear{{McGreer}, {Mesinger}  \&
  {D'Odorico}}{{McGreer} et~al.}{2015}]{2015MNRAS.447..499M}
{McGreer} I.~D.,  {Mesinger} A.,   {D'Odorico} V.,  2015, \mn@doi [\mnras]
  {10.1093/mnras/stu2449}, \href
  {http://adsabs.harvard.edu/abs/2015MNRAS.447..499M} {447, 499}

\bibitem[\protect\citeauthoryear{{McLeod}, {McLure}, {Dunlop}, {Robertson},
  {Ellis}  \& {Targett}}{{McLeod} et~al.}{2014}]{2014arXiv1412.1472M}
{McLeod} D.~J.,  {McLure} R.~J.,  {Dunlop} J.~S.,  {Robertson} B.~E.,  {Ellis}
  R.~S.,   {Targett} T.~T.,  2014, arXiv:1412.1472, \href
  {http://adsabs.harvard.edu/abs/2014arXiv1412.1472M} {}

\bibitem[\protect\citeauthoryear{{McQuinn}}{{McQuinn}}{2016}]{2016ARA&A..54..313M}
{McQuinn} M.,  2016, \mn@doi [\araa] {10.1146/annurev-astro-082214-122355},
  \href {http://adsabs.harvard.edu/abs/2016ARA%26A..54..313M} {54, 313}

\bibitem[\protect\citeauthoryear{{McQuinn}, {Lidz}, {Zaldarriaga}, {Hernquist},
  {Hopkins}, {Dutta}  \& {Faucher-Gigu{\`e}re}}{{McQuinn}
  et~al.}{2009}]{2009ApJ...694..842M}
{McQuinn} M.,  {Lidz} A.,  {Zaldarriaga} M.,  {Hernquist} L.,  {Hopkins} P.~F.,
   {Dutta} S.,   {Faucher-Gigu{\`e}re} C.-A.,  2009, \mn@doi [\apj]
  {10.1088/0004-637X/694/2/842}, \href
  {http://adsabs.harvard.edu/abs/2009ApJ...694..842M} {694, 842}

\bibitem[\protect\citeauthoryear{{Mesinger}, {Aykutalp}, {Vanzella},
  {Pentericci}, {Ferrara}  \& {Dijkstra}}{{Mesinger}
  et~al.}{2015}]{2015MNRAS.446..566M}
{Mesinger} A.,  {Aykutalp} A.,  {Vanzella} E.,  {Pentericci} L.,  {Ferrara} A.,
    {Dijkstra} M.,  2015, \mn@doi [\mnras] {10.1093/mnras/stu2089}, \href
  {http://adsabs.harvard.edu/abs/2015MNRAS.446..566M} {446, 566}

\bibitem[\protect\citeauthoryear{{Miralda-Escud{\'e}} \&
  {Rees}}{{Miralda-Escud{\'e}} \& {Rees}}{1994}]{1994MNRAS.266..343M}
{Miralda-Escud{\'e}} J.,  {Rees} M.~J.,  1994, \mnras, \href
  {http://adsabs.harvard.edu/abs/1994MNRAS.266..343M} {266, 343}

\bibitem[\protect\citeauthoryear{{Miralda-Escud{\'e}}, {Haehnelt}  \&
  {Rees}}{{Miralda-Escud{\'e}} et~al.}{2000}]{2000ApJ...530....1M}
{Miralda-Escud{\'e}} J.,  {Haehnelt} M.,   {Rees} M.~J.,  2000, \mn@doi [\apj]
  {10.1086/308330}, \href {http://adsabs.harvard.edu/abs/2000ApJ...530....1M}
  {530, 1}

\bibitem[\protect\citeauthoryear{{Mitra}, {Choudhury}  \& {Ferrara}}{{Mitra}
  et~al.}{2011}]{mitra1}
{Mitra} S.,  {Choudhury} T.~R.,   {Ferrara} A.,  2011, \mn@doi [\mnras]
  {10.1111/j.1365-2966.2011.18234.x}, \href
  {http://adsabs.harvard.edu/abs/2011MNRAS.413.1569M} {413, 1569}

\bibitem[\protect\citeauthoryear{{Mitra}, {Choudhury}  \& {Ferrara}}{{Mitra}
  et~al.}{2012}]{mitra2}
{Mitra} S.,  {Choudhury} T.~R.,   {Ferrara} A.,  2012, \mn@doi [\mnras]
  {10.1111/j.1365-2966.2011.19804.x}, \href
  {http://adsabs.harvard.edu/abs/2012MNRAS.419.1480M} {419, 1480}

\bibitem[\protect\citeauthoryear{{Mitra}, {Ferrara}  \& {Choudhury}}{{Mitra}
  et~al.}{2013}]{mitra3}
{Mitra} S.,  {Ferrara} A.,   {Choudhury} T.~R.,  2013, \mn@doi [\mnras]
  {10.1093/mnrasl/sls001}, \href
  {http://adsabs.harvard.edu/abs/2013MNRAS.428L...1M} {428, L1}

\bibitem[\protect\citeauthoryear{{Mitra}, {Choudhury}  \& {Ferrara}}{{Mitra}
  et~al.}{2015}]{mitra4}
{Mitra} S.,  {Choudhury} T.~R.,   {Ferrara} A.,  2015, \mn@doi [\mnras]
  {10.1093/mnrasl/slv134}, \href
  {http://adsabs.harvard.edu/abs/2015MNRAS.454L..76M} {454, L76}

\bibitem[\protect\citeauthoryear{{Oesch} et~al.}{{Oesch}
  et~al.}{2012}]{2012ApJ...745..110O}
{Oesch} P.~A.,  et~al., 2012, \mn@doi [\apj] {10.1088/0004-637X/745/2/110},
  \href {http://adsabs.harvard.edu/abs/2012ApJ...745..110O} {745, 110}

\bibitem[\protect\citeauthoryear{{Oesch} et~al.}{{Oesch}
  et~al.}{2014}]{2014ApJ...786..108O}
{Oesch} P.~A.,  et~al., 2014, \mn@doi [\apj] {10.1088/0004-637X/786/2/108},
  \href {http://adsabs.harvard.edu/abs/2014ApJ...786..108O} {786, 108}

\bibitem[\protect\citeauthoryear{{Okamoto}, {Gao}  \& {Theuns}}{{Okamoto}
  et~al.}{2008}]{2008MNRAS.390..920O}
{Okamoto} T.,  {Gao} L.,   {Theuns} T.,  2008, \mn@doi [\mnras]
  {10.1111/j.1365-2966.2008.13830.x}, \href
  {http://adsabs.harvard.edu/abs/2008MNRAS.390..920O} {390, 920}

\bibitem[\protect\citeauthoryear{{Ota} et~al.}{{Ota}
  et~al.}{2008}]{2008ApJ...677...12O}
{Ota} K.,  et~al., 2008, \mn@doi [\apj] {10.1086/529006}, \href
  {http://adsabs.harvard.edu/abs/2008ApJ...677...12O} {677, 12}

\bibitem[\protect\citeauthoryear{{Ouchi} et~al.}{{Ouchi}
  et~al.}{2010}]{2010ApJ...723..869O}
{Ouchi} M.,  et~al., 2010, \mn@doi [\apj] {10.1088/0004-637X/723/1/869}, \href
  {http://adsabs.harvard.edu/abs/2010ApJ...723..869O} {723, 869}

\bibitem[\protect\citeauthoryear{{Padmanabhan}, {Choudhury}  \&
  {Srianand}}{{Padmanabhan} et~al.}{2014}]{2014MNRAS.443.3761P}
{Padmanabhan} H.,  {Choudhury} T.~R.,   {Srianand} R.,  2014, \mn@doi [\mnras]
  {10.1093/mnras/stu1433}, \href
  {http://adsabs.harvard.edu/abs/2014MNRAS.443.3761P} {443, 3761}

\bibitem[\protect\citeauthoryear{{Planck Collaboration} et~al.,}{{Planck
  Collaboration} et~al.}{2014}]{2014A&A...571A..16P}
{Planck Collaboration} et~al., 2014, \mn@doi [\aap]
  {10.1051/0004-6361/201321591}, \href
  {http://adsabs.harvard.edu/abs/2014A%26A...571A..16P} {571, A16}

\bibitem[\protect\citeauthoryear{{Planck Collaboration} et~al.,}{{Planck
  Collaboration} et~al.}{2016a}]{2016A&A...594A..13P}
{Planck Collaboration} et~al., 2016a, \mn@doi [\aap]
  {10.1051/0004-6361/201525830}, \href
  {http://adsabs.harvard.edu/abs/2016A%26A...594A..13P} {594, A13}

\bibitem[\protect\citeauthoryear{{Planck Collaboration} et~al.,}{{Planck
  Collaboration} et~al.}{2016b}]{2016A&A...596A.107P}
{Planck Collaboration} et~al., 2016b, \mn@doi [\aap]
  {10.1051/0004-6361/201628890}, \href
  {http://adsabs.harvard.edu/abs/2016A%26A...596A.107P} {596, A107}

\bibitem[\protect\citeauthoryear{{Planck Collaboration} et~al.,}{{Planck
  Collaboration} et~al.}{2016c}]{2016A&A...596A.108P}
{Planck Collaboration} et~al., 2016c, \mn@doi [\aap]
  {10.1051/0004-6361/201628897}, \href
  {http://adsabs.harvard.edu/abs/2016A%26A...596A.108P} {596, A108}

\bibitem[\protect\citeauthoryear{{Price}, {Trac}  \& {Cen}}{{Price}
  et~al.}{2016}]{2016arXiv160503970P}
{Price} L.~C.,  {Trac} H.,   {Cen} R.,  2016, preprint, \href
  {http://adsabs.harvard.edu/abs/2016arXiv160503970P} {} (\mn@eprint {arXiv}
  {1605.03970})

\bibitem[\protect\citeauthoryear{{Prochaska}, {O'Meara}  \&
  {Worseck}}{{Prochaska} et~al.}{2010}]{2010ApJ...718..392P}
{Prochaska} J.~X.,  {O'Meara} J.~M.,   {Worseck} G.,  2010, \mn@doi [\apj]
  {10.1088/0004-637X/718/1/392}, \href
  {http://adsabs.harvard.edu/abs/2010ApJ...718..392P} {718, 392}

\bibitem[\protect\citeauthoryear{{Puchwein}, {Pfrommer}, {Springel},
  {Broderick}  \& {Chang}}{{Puchwein} et~al.}{2012}]{2012MNRAS.423..149P}
{Puchwein} E.,  {Pfrommer} C.,  {Springel} V.,  {Broderick} A.~E.,   {Chang}
  P.,  2012, \mn@doi [\mnras] {10.1111/j.1365-2966.2012.20738.x}, \href
  {http://adsabs.harvard.edu/abs/2012MNRAS.423..149P} {423, 149}

\bibitem[\protect\citeauthoryear{{Raskutti}, {Bolton}, {Wyithe}  \&
  {Becker}}{{Raskutti} et~al.}{2012}]{2012MNRAS.421.1969R}
{Raskutti} S.,  {Bolton} J.~S.,  {Wyithe} J.~S.~B.,   {Becker} G.~D.,  2012,
  \mn@doi [\mnras] {10.1111/j.1365-2966.2011.20401.x}, \href
  {http://adsabs.harvard.edu/abs/2012MNRAS.421.1969R} {421, 1969}

\bibitem[\protect\citeauthoryear{{Richards} et~al.,}{{Richards}
  et~al.}{2006}]{2006ApJS..166..470R}
{Richards} G.~T.,  et~al., 2006, \mn@doi [\apjs] {10.1086/506525}, \href
  {http://adsabs.harvard.edu/abs/2006ApJS..166..470R} {166, 470}

\bibitem[\protect\citeauthoryear{{Robertson}, {Ellis}, {Furlanetto}  \&
  {Dunlop}}{{Robertson} et~al.}{2015}]{2015ApJ...802L..19R}
{Robertson} B.~E.,  {Ellis} R.~S.,  {Furlanetto} S.~R.,   {Dunlop} J.~S.,
  2015, \mn@doi [\apjl] {10.1088/2041-8205/802/2/L19}, \href
  {http://adsabs.harvard.edu/abs/2015ApJ...802L..19R} {802, L19}

\bibitem[\protect\citeauthoryear{{Samui}, {Srianand}  \& {Subramanian}}{{Samui}
  et~al.}{2007}]{2007MNRAS.377..285S}
{Samui} S.,  {Srianand} R.,   {Subramanian} K.,  2007, \mn@doi [\mnras]
  {10.1111/j.1365-2966.2007.11603.x}, \href
  {http://adsabs.harvard.edu/abs/2007MNRAS.377..285S} {377, 285}

\bibitem[\protect\citeauthoryear{{Schenker}, {Ellis}, {Konidaris}  \&
  {Stark}}{{Schenker} et~al.}{2014}]{2014ApJ...795...20S}
{Schenker} M.~A.,  {Ellis} R.~S.,  {Konidaris} N.~P.,   {Stark} D.~P.,  2014,
  \mn@doi [\apj] {10.1088/0004-637X/795/1/20}, \href
  {http://adsabs.harvard.edu/abs/2014ApJ...795...20S} {795, 20}

\bibitem[\protect\citeauthoryear{{Schirber} \& {Bullock}}{{Schirber} \&
  {Bullock}}{2003}]{2003ApJ...584..110S}
{Schirber} M.,  {Bullock} J.~S.,  2003, \mn@doi [\apj] {10.1086/345662}, \href
  {http://adsabs.harvard.edu/abs/2003ApJ...584..110S} {584, 110}

\bibitem[\protect\citeauthoryear{{Schroeder}, {Mesinger}  \&
  {Haiman}}{{Schroeder} et~al.}{2013}]{2013MNRAS.428.3058S}
{Schroeder} J.,  {Mesinger} A.,   {Haiman} Z.,  2013, \mn@doi [\mnras]
  {10.1093/mnras/sts253}, \href
  {http://adsabs.harvard.edu/abs/2013MNRAS.428.3058S} {428, 3058}

\bibitem[\protect\citeauthoryear{{Sharma}, {Theuns}, {Frenk}, {Bower}, {Crain},
  {Schaller}  \& {Schaye}}{{Sharma} et~al.}{2016}]{2016MNRAS.458L..94S}
{Sharma} M.,  {Theuns} T.,  {Frenk} C.,  {Bower} R.,  {Crain} R.,  {Schaller}
  M.,   {Schaye} J.,  2016, \mn@doi [\mnras] {10.1093/mnrasl/slw021}, \href
  {http://adsabs.harvard.edu/abs/2016MNRAS.458L..94S} {458, L94}

\bibitem[\protect\citeauthoryear{{Sobacchi} \& {Mesinger}}{{Sobacchi} \&
  {Mesinger}}{2013}]{2013MNRAS.432.3340S}
{Sobacchi} E.,  {Mesinger} A.,  2013, \mn@doi [\mnras] {10.1093/mnras/stt693},
  \href {http://adsabs.harvard.edu/abs/2013MNRAS.432.3340S} {432, 3340}

\bibitem[\protect\citeauthoryear{{Songaila} \& {Cowie}}{{Songaila} \&
  {Cowie}}{2010}]{2010ApJ...721.1448S}
{Songaila} A.,  {Cowie} L.~L.,  2010, \mn@doi [\apj]
  {10.1088/0004-637X/721/2/1448}, \href
  {http://adsabs.harvard.edu/abs/2010ApJ...721.1448S} {721, 1448}

\bibitem[\protect\citeauthoryear{{Totani} et~al.}{{Totani}
  et~al.}{2006}]{2006PASJ...58..485T}
{Totani} T.,  et~al., 2006, \pasj, \href
  {http://adsabs.harvard.edu/abs/2006PASJ...58..485T} {58, 485}

\bibitem[\protect\citeauthoryear{{Trac}, {Cen}  \& {Loeb}}{{Trac}
  et~al.}{2008}]{2008ApJ...689L..81T}
{Trac} H.,  {Cen} R.,   {Loeb} A.,  2008, \mn@doi [\apjl] {10.1086/595678},
  \href {http://adsabs.harvard.edu/abs/2008ApJ...689L..81T} {689, L81}

\bibitem[\protect\citeauthoryear{{Upton Sanderbeck}, {D'Aloisio}  \&
  {McQuinn}}{{Upton Sanderbeck} et~al.}{2016}]{2016MNRAS.460.1885U}
{Upton Sanderbeck} P.~R.,  {D'Aloisio} A.,   {McQuinn} M.~J.,  2016, \mn@doi
  [\mnras] {10.1093/mnras/stw1117}, \href
  {http://adsabs.harvard.edu/abs/2016MNRAS.460.1885U} {460, 1885}

\bibitem[\protect\citeauthoryear{{White}, {Becker}, {Fan}  \&
  {Strauss}}{{White} et~al.}{2003}]{2003AJ....126....1W}
{White} R.~L.,  {Becker} R.~H.,  {Fan} X.,   {Strauss} M.~A.,  2003, \mn@doi
  [\aj] {10.1086/375547}, \href
  {http://adsabs.harvard.edu/abs/2003AJ....126....1W} {126, 1}

\bibitem[\protect\citeauthoryear{{Worseck} et~al.}{{Worseck}
  et~al.}{2014}]{2014MNRAS.445.1745W}
{Worseck} G.,  et~al., 2014, \mn@doi [\mnras] {10.1093/mnras/stu1827}, \href
  {http://adsabs.harvard.edu/abs/2014MNRAS.445.1745W} {445, 1745}

\bibitem[\protect\citeauthoryear{{Worseck}, {Prochaska}, {Hennawi}  \&
  {McQuinn}}{{Worseck} et~al.}{2016}]{2016ApJ...825..144W}
{Worseck} G.,  {Prochaska} J.~X.,  {Hennawi} J.~F.,   {McQuinn} M.,  2016,
  \mn@doi [\apj] {10.3847/0004-637X/825/2/144}, \href
  {http://adsabs.harvard.edu/abs/2016ApJ...825..144W} {825, 144}

\bibitem[\protect\citeauthoryear{{Wyithe} \& {Loeb}}{{Wyithe} \&
  {Loeb}}{2003}]{2003ApJ...586..693W}
{Wyithe} J.~S.~B.,  {Loeb} A.,  2003, \mn@doi [\apj] {10.1086/367721}, \href
  {http://adsabs.harvard.edu/abs/2003ApJ...586..693W} {586, 693}

\bibitem[\protect\citeauthoryear{{Wyithe} \& {Loeb}}{{Wyithe} \&
  {Loeb}}{2005}]{2005ApJ...625....1W}
{Wyithe} J.~S.~B.,  {Loeb} A.,  2005, \mn@doi [\apj] {10.1086/429529}, \href
  {http://adsabs.harvard.edu/abs/2005ApJ...625....1W} {625, 1}

\bibitem[\protect\citeauthoryear{{Yoshiura}, {Hasegawa}, {Ichiki}, {Tashiro},
  {Shimabukuro}  \& {Takahashi}}{{Yoshiura} et~al.}{2016}]{2016arXiv160204407Y}
{Yoshiura} S.,  {Hasegawa} K.,  {Ichiki} K.,  {Tashiro} H.,  {Shimabukuro} H.,
   {Takahashi} K.,  2016, preprint, \href
  {http://adsabs.harvard.edu/abs/2016arXiv160204407Y} {} (\mn@eprint {arXiv}
  {1602.04407})

\bibitem[\protect\citeauthoryear{{Zaldarriaga}, {Hui}  \&
  {Tegmark}}{{Zaldarriaga} et~al.}{2001}]{2001ApJ...557..519Z}
{Zaldarriaga} M.,  {Hui} L.,   {Tegmark} M.,  2001, \mn@doi [\apj]
  {10.1086/321652}, \href {http://adsabs.harvard.edu/abs/2001ApJ...557..519Z}
  {557, 519}

\makeatother
\end{thebibliography}
\bibliographystyle{mnras}

\end{document}